\documentclass[aps,prb,superscriptaddress,reprint,floatfix]{revtex4-2}
\usepackage[utf8]{inputenc}
\usepackage{graphicx,dcolumn,bm,amsmath,amssymb,euscript}

\usepackage{color, colortbl}   %% in text, use as \textcolor{green}{text that shows up green}
\usepackage[bookmarksopen,bookmarksnumbered,colorlinks,citecolor=red,urlcolor=blue,filecolor=green]{hyperref}

\begin{document}

\title{Step- and terrace-resolved crystal truncation rod scattering from vicinal surfaces under coherent heteroepitaxy}

\author{Junlin Wu}
	\affiliation{State Key Laboratory of Artificial Microstructure and Mesoscopic Physics, School of Physics, Peking University, Beijing 100871, P. R. China}
    
\author{Erqi Xu}
	\affiliation{State Key Laboratory of Artificial Microstructure and Mesoscopic Physics, School of Physics, Peking University, Beijing 100871, P. R. China}
    
\author{Qihui Lin}
	\affiliation{State Key Laboratory of Artificial Microstructure and Mesoscopic Physics, School of Physics, Peking University, Beijing 100871, P. R. China}
\author{Jiaqing Yue}
	\affiliation{State Key Laboratory of Artificial Microstructure and Mesoscopic Physics, School of Physics, Peking University, Beijing 100871, P. R. China}

\author{Jiale Wang}
	\affiliation{State Key Laboratory of Artificial Microstructure and Mesoscopic Physics, School of Physics, Peking University, Beijing 100871, P. R. China}

\author{Zihao Xu}
	\affiliation{State Key Laboratory of Artificial Microstructure and Mesoscopic Physics, School of Physics, Peking University, Beijing 100871, P. R. China}

\author{Guangxu Ju}
    \email[correspondence to: ]{gxju@pku.edu.cn}
	\affiliation{State Key Laboratory of Artificial Microstructure and Mesoscopic Physics, School of Physics, Peking University, Beijing 100871, P. R. China}

%\date{revision %\VERSION  : \today}
%\date{revision \VERSION  : \today}

\begin{abstract}
We develop a general theory of crystal truncation rod (CTR) scattering from vicinal surfaces with a coherently strained heteroepitaxial film. The formalism incorporates film-induced interference fringes, full elastic lattice distortion, terrace ordering, surface reconstruction, and real-time growth evolution within a unified description. Comparison between Nagai model and elasticity-based model shows that the lattice tilt is nearly identical in the two approaches, whereas the elasticity-based model predicts an additional triclinic deformation arising from shear strain. This deformation has little effect on specular CTRs but strongly modifies non-specular rods, making them a sensitive probe of the full elastic state of coherent epitaxial films. We further show that the characteristic sensitivity of vicinal CTRs to terrace ordering, surface reconstruction, and terrace-resolved compositional modification remains robust in the presence of a coherent film. Representative calculations for InGaN/GaN demonstrate that the framework enables quantitative interpretation of both static and real-time CTR measurements and provides access to step- and terrace-resolved structural and kinetic information during heteroepitaxial growth.
\end{abstract}

\maketitle
\section{Introduction}

Surface X-ray scattering provides a powerful approach for determining the atomic-scale structure of epitaxial films through measurements of the intensity distribution along crystal truncation rods (CTRs)\cite{2006_Fong_ARMS36_431,2020_Disa_AdvMarInterf7_1901772}. 
These rods arise from the truncation of the bulk crystal lattice at the surface and extend normal to the surface from each Bragg peak\cite{1986_Robinson_PRB33_3830,petach2017crystal,wang2025determination}. 
CTR profiles are sensitive to multiple structural parameters, including lattice strain, film thickness, surface termination, and atomic-scale roughness\cite{1997_Thompson_APL71_3516,2004_Fenter_JApplCryst37_977,2017_Petach_PRB95_184104,2021_Ju_NatCommun12_1721}. 
In particular, the positions of Bragg peaks along the CTR provide direct information on the out-of-plane lattice parameters, while interference fringes enable precise determination of film thickness\cite{1997_Thompson_APL71_3516,ju2014situ}. 
Under coherent heteroepitaxial growth conditions, the overlap between film and substrate CTRs further encodes information about the interface structure\cite{1988_Robinson_PRB38_3632,2008_Kaganer_PRB77_125325,2009_LeBeau_APL95_142905,ju2013situ}.

Analyses of CTRs are often formulated for surfaces aligned with crystallographic planes (zero off-cut), where rods from different Bragg peaks share identical in-plane components and therefore overlap\cite{2021_Ju_PRB_CTR,1997_Thompson_APL71_3516}. 
In practice, however, epitaxial growth is frequently carried out on vicinal substrates, where the surface is intentionally misoriented by a small angle from a low-index crystallographic plane. 
Such off-cut substrates promote step-flow growth by reducing terrace widths below the adatom diffusion length, leading to surfaces composed of periodic terraces separated by atomic steps\cite{1993_Tsao_MatFundMBE_Ch6,sarzynski2012influence}.

For vicinal surfaces, the CTRs are tilted relative to the crystal lattice, and rods from different Bragg peaks no longer coincide\cite{1995_Held_PRB51_7262,2002_Trainor_JApplCryst35_696,2007_Wollschlager_PRB75_245439,2017_Petach_PRB95_184104,2021_Ju_PRB_CTR}. 
This geometric modification introduces additional structure into the CTR profiles and enables access to step-resolved information such as terrace widths and step-edge configurations. 
CTR measurements on vicinal GaN(0001) surfaces have revealed the relative widths of $\alpha$ and $\beta$ terraces and their distinct kinetic behavior\cite{2021_Ju_NatCommun12_1721,2020_Ju_PRB_BCF}.

\begin{figure*}[t!]
\includegraphics[width=0.9\linewidth]{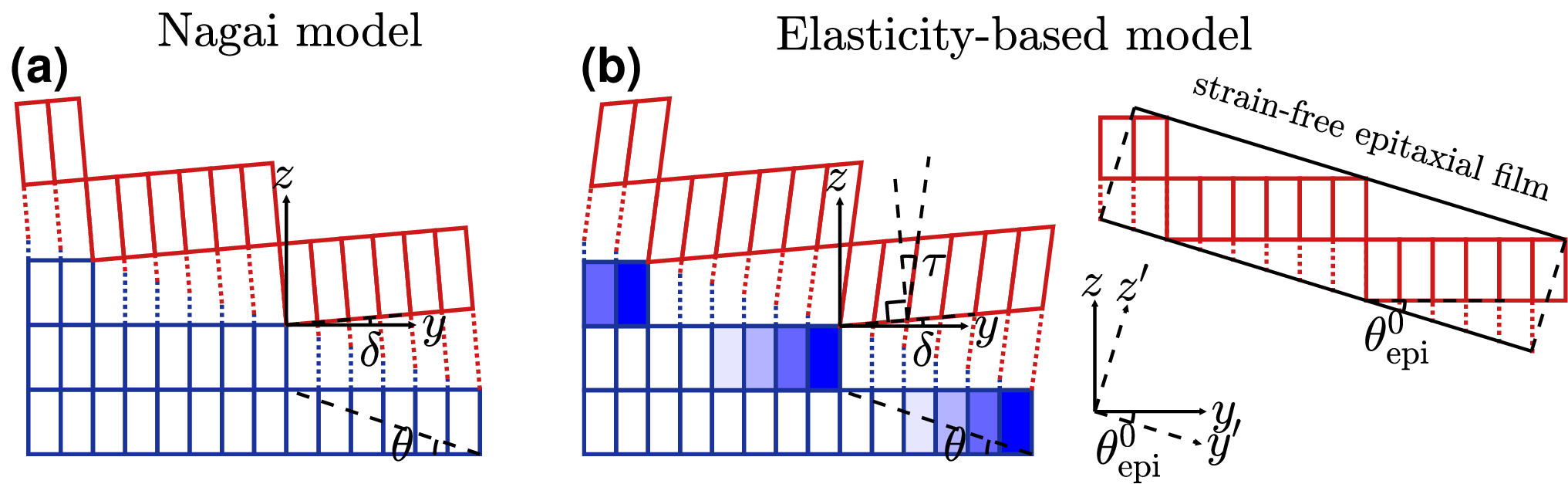}
\caption{Sketch of (a) the Nagai model and (b) the elasticity-based model for a fully strained epitaxial film on a vicinal substrate. (a) The substrate coordinate axes are defined by $z \parallel [001]$ and $y \parallel [010]$. The substrate off-cut angle is $\theta$, and $\delta$ denotes the lattice tilt of the epilayer relative to substrate. (b) In the elasticity-based model, $\tau$ characterizes the triclinic deformation, corresponding to a shear-induced deviation of the strained lattice from an orthogonal geometry. The quantity $\theta_\mathrm{epi}^0$ is the off-cut angle of the strain-free epilayer. The reference coordinate system $(y', z')$ is defined with $z'$ normal to the surface of the strain-free epilayer. The strained epilayer lattice is then obtained from this reference configuration through elastic deformation and rigid-body rotation, as described in Eq.~(\ref{eq:s_latt}). }
\label{fig:InGaNonGaN}
\end{figure*}

Despite these advances, a general CTR formalism that simultaneously captures vicinal geometry, coherent film interference, full elastic distortion, and terrace-resolved surface structure is still lacking. 
In heteroepitaxial systems, lattice mismatch introduces additional geometric and elastic effects beyond those in homoepitaxy. Even when the film surface remains parallel to the substrate, the film lattice is rotated relative to the substrate due to the difference in off-cut angles, as described by the Nagai model and its extensions\cite{1974_Nagai_JAP45_3789,1986_Macrander_JAP59_442,1988_Neumann_JAP64_3024,2010_Krysko_PSS_RRL4_142,2017_Suzuki_JJAP56_08MA06}. However, this description accounts only for normal strain and rotation, and does not capture the full elastic response of a coherently strained film on a vicinal substrate. Beyond pure rotation, elasticity theory predicts a triclinic deformation involving both lattice vectors and interaxial angles\cite{2013_Krysko_JAP114_113512,2021_Liao_JAP129_025105}, which significantly modifies the lattice geometry (Fig.~\ref{fig:InGaNonGaN}). Accurate CTR analysis therefore requires an elasticity-based description that accounts for the full strain tensor and resulting lattice geometry.

Alloying introduces additional complexity through composition-dependent surface processes. For example, indium incorporation in InGaN exhibits surface segregation and step-dependent behavior\cite{2006_Jiang_APL89_161915}, leading to local compositional variations at step edges that are not captured by existing CTR formalisms for uniform films.

Here we develop a general theoretical framework for CTR scattering from vicinal surfaces with a coherent heteroepitaxial film. 
Building on previous formulations \cite{2021_Ju_PRB_CTR}, the present approach incorporates full elastic lattice distortion, film-induced interference fringes, terrace-resolved structural variations, and surface reconstructions within a unified description.  
To account for alloy effects, sub-monolayer surface regions with compositions distinct from the underlying film are introduced, enabling modeling of step-edge segregation.

We apply the framework to coherent InGaN films on vicinal GaN(0001) and calculate the resulting CTR intensity distributions and their evolution during growth. 
The results provide a basis for quantitative interpretation of static and real-time CTR measurements and enable extraction of step-resolved structural and kinetic information under realistic growth conditions.

\section{\label{sec:II}Strain-induced lattice tilting and deformation in coherent heteroepitaxy on vicinal surfaces}

For coherent heteroepitaxy on a vicinal surface, lattice mismatch induces not only normal strain but also a tilt of the epilayer lattice relative to the substrate. In addition to this geometric tilt, the off-cut geometry can give rise to further elastic distortions associated with shear strain. 

In this section, we analyze these effects using two complementary models: the classical Nagai model\cite{1974_Nagai_JAP45_3789}, which describes lattice tilting as a rigid-body rotation of a normally strained lattice, and an elasticity-based model, which captures the full strain tensor and the resulting triclinic deformation. A schematic comparison of the two descriptions is shown in Fig.~\ref{fig:InGaNonGaN}.

Throughout this paper, the subscript “epi” denotes the epitaxial film, “rec” the surface reconstruction, and the superscript “0” the stress-free state. Quantities without subscripts, or with the subscript “bulk”, refer to the bulk substrate. We adopt an orthorhombic lattice representation in a Cartesian coordinate system $(x,y,z)$ defined by the mutually orthogonal substrate lattice vectors $\mathbf{a}, \mathbf{b}, \mathbf{c}$, with magnitudes $a$, $b$, and $c$, respectively. To describe the vicinal surface, we consider a periodic array of steps parallel to the $x$ direction, with a step height of one unit cell and a terrace width of $M$ lattice spacings along the $y$ (step-down) direction (Fig.~\ref{fig:InGaNonGaN}).

\subsection{Classical Nagai model: lattice tilting from mismatch}
The geometry of the Nagai model is shown in Fig.~\ref{fig:InGaNonGaN}(a). In this model, the epilayer is assumed to undergo no shear deformation, and lattice mismatch is accommodated solely through normal strain and a rotation of the film lattice. The tilt angle is given by
\begin{equation}
\delta = \theta_\mathrm{epi} - \theta,
\end{equation}
where $\theta$ is the substrate off-cut angle and $\theta_\mathrm{epi}$ is that of the strained epilayer. For a vicinal surface with one-unit-cell step height and terrace width $Mb$, $\tan\theta = c/(Mb)$ and $\tan\theta_\mathrm{epi} = c_\mathrm{epi}/(Mb_\mathrm{epi})$. 

Lattice matching requires
\begin{equation}
    (Mb)^2 + c^2 = (Mb_\mathrm{epi})^2 + c_\mathrm{epi}^2,
    \label{eq:latt_match_Nagai}
\end{equation}
while the strain is given by in-plane Hooke's law
\begin{equation}
    \frac{c_\mathrm{epi}-c_\mathrm{epi}^0}{c_\mathrm{epi}^0}=-\frac{c_{13}}{c_{33}} 
    \frac{a_\mathrm{epi}-a_\mathrm{epi}^0}{a_\mathrm{epi}^0} -\frac{c_{23}}{c_{33}} 
    \frac{b_\mathrm{epi}-b_\mathrm{epi}^0}{b_\mathrm{epi}^0},
    \label{eq:strain_Nagai}
\end{equation}
where $c_{ij}$ are elastic constants. For a fully strained epilayer,  $a_\mathrm{epi}=a$ and $b_{\mathrm{epi}}=b+\mathcal{O}[(c-c_{\mathrm{epi}})/M^2]$ [from Eq.~(\ref{eq:latt_match_Nagai})]. The correction to $b_{\mathrm{epi}}$ is negligible for large $M$. 

For materials with in-plane isotropy (e.g., hexagonal systems), the two in-plane strain components are equivalent: $(a-a_\mathrm{epi}^0)/{a_\mathrm{epi}^0} = (b-b_\mathrm{epi}^0)/{b_\mathrm{epi}^0}$, and the elastic constants satisfy $c_{13}=c_{23}$. The strain expression, therefore, reduces to
\begin{equation}
    \frac{c_\mathrm{epi}-c_\mathrm{epi}^0}{c_\mathrm{epi}^0}=-\frac{2c_{13}}{c_{33}} 
    \frac{a-a_\mathrm{epi}^0}{a_\mathrm{epi}^0}.
    \label{eq:c_epi_N}
\end{equation}
Together with Eq.~(\ref{eq:latt_match_Nagai}), the two unknowns $b_\mathrm{epi},c_\mathrm{epi}$ are determined, and from these one obtains $\theta_\mathrm{epi}$ and hence $\delta$. 

The Nagai model provides a useful description of lattice tilting in coherent epitaxial films, capturing the geometric origin of the tilt arising from lattice mismatch on vicinal surfaces. However, this model treats lattice distortion as normal strain with a rigid-body rotation and does not account for the full elastic response of a strained film. In particular, the off-cut geometry introduces additional triclinic deformation, associated with shear components of the strain tensor. It modifies the crystal geometry and affects the interpretation of CTR measurements, especially for non-specular rods. To capture these effects, we adopt a refined crystal model based on elasticity theory.

\subsection{\label{sb:IIB}Elasticity-based model: triclinic deformation and refined geometry}

In contrast to the Nagai model, additional degrees of freedom induced by shear strain lead to a more general lattice distortion beyond a rigid-body rotation. The epilayer undergoes a triclinic deformation characterized by the tilt angle $\tau$, as illustrated in Fig.~\ref{fig:InGaNonGaN}(b). This effect was first reported experimentally in the InGaN/GaN system by Krysko \textit{et al.}\cite{2013_Krysko_JAP114_113512}. The strain is calculated within the framework of linear elasticity, following the formalism of Romanov \textit{et al.}\cite{2006_romanov_JAP100_023522}. In the present work, we extend this approach by explicitly constructing the strained lattice vectors required for CTR calculations.

To describe the epilayer geometry, we introduce two right-handed coordinate systems, as shown in Fig.~\ref{fig:InGaNonGaN}(b). The crystallographic system $(x,y,z)$ is aligned with the substrate lattice axes: $x \parallel [100]$, $y \parallel [010]$, and $z \parallel [001]$. The surface-adapted system $(x',y',z')$ is defined for the strain-free epilayer, with $x'y'$ plane parallel to the vicinal surface, $z'$ along the surface normal, $x' \parallel [100]$, and $y'$ along the step-down direction. The two systems are related by a rotation about the common $x$ axis by $\theta_\mathrm{epi}^0$.

The stress and strain are related by Hooke's law
\begin{equation}
\sigma_{ij}=C_{ijkl}\epsilon_{kl},
\label{eq:Hooke_law} 
\end{equation}
where $C_{ijkl}$ are epilayer elastic constants in the crystallographic frame. The strain and stress tensors each contain six independent components, given that they are symmetric. Thus, Hooke's law is conveniently expressed in Voigt notation [Eq.~(\ref{aeq:hook})].  

The boundary conditions are specified in the $x',y',z'$  coordinate system. The in-plane strain is assumed to be biaxial with no shear component, consistent with straight step edges: 
\begin{equation}
\epsilon_{x^{\prime} x^\prime}=\epsilon_{m1};\quad \epsilon_{y^{\prime} y^{\prime}}=\epsilon_{m2};\quad \epsilon_{x^{\prime} y^{\prime}}=0,
\label{eq:inplane} 
\end{equation}
where $\epsilon_{m1},\epsilon_{m2}$ are determined from lattice matching conditions. Along the step-edge direction,  
\begin{equation}
\epsilon_{m1}=\frac{a}{a_\mathrm{epi}^0}-1,
\label{eq:x_match}
\end{equation}
and along the step-down direction, 
\begin{equation}
\epsilon_{m2}=\frac{\sqrt{(Mb)^2 + c^2}}{\sqrt{(Mb_\mathrm{epi}^0)^2 + (c_\mathrm{epi}^0)^2}}-1.
\label{eq:step-down_match}
\end{equation}

The surface is assumed to be stress-free, giving
\begin{equation}
\sigma_{x^{\prime} z^{\prime}}=0;\quad \sigma_{y^{\prime} z^{\prime}}=0;\quad \sigma_{z^{\prime} z^{\prime}}=0. 
\label{eq:free_plane} 
\end{equation}

With Eqs.~(\ref{eq:Hooke_law}), (\ref{eq:inplane}), and (\ref{eq:free_plane}) — comprising twelve equations (six from Hooke's law and six from boundary conditions) for twelve unknowns (six strain and six stress components) — the strain tensor $\boldsymbol{\epsilon}$ is fully determined, since the tensors in the surface-adapted coordinate system and the crystallographic system are related by a coordinate transformation [Eq.~(\ref{aeq:coor_transf})].

The strained lattice vectors are then given by
\begin{equation}
\mathbf{v}_\mathrm{epi}=\boldsymbol{T}_x(\gamma)(\boldsymbol{I}+\boldsymbol{\epsilon}) \mathbf{v}_\mathrm{epi}^0.
\label{eq:s_latt} 
\end{equation}
Here, $\mathbf{v}_\mathrm{epi}^0$ denotes an arbitrary lattice vector in the reference configuration of the strain-free epilayer. In particular, the strained lattice vectors $\mathbf{a}_\mathrm{epi}$, $\mathbf{b}_\mathrm{epi}$, and $\mathbf{c}_\mathrm{epi}$ are obtained by applying the same transformation to $\mathbf{a}_\mathrm{epi}^0$, $\mathbf{b}_\mathrm{epi}^0$, and $\mathbf{c}_\mathrm{epi}^0$, respectively. The operator $(\boldsymbol{I}+\boldsymbol{\epsilon})$ describes the deformation of the epilayer lattice due to elastic strain, while the rotation $\boldsymbol{T}_x(\gamma)$ ensures that the bottom surface of the strained epilayer remains parallel to the substrate surface [see Appendix~Eq.~(\ref{aeq:Rx})], which is necessary because the elastic strain tensor does not include rigid-body rotation. Thus, the final lattice vectors incorporate the full elastic constraint imposed by the substrate. The lattice tilt $\delta$ and triclinic tilt $\tau$ are then extracted from the orientation and interaxial angles of the strained lattice.

\begin{figure}
\includegraphics[width=0.9\linewidth]{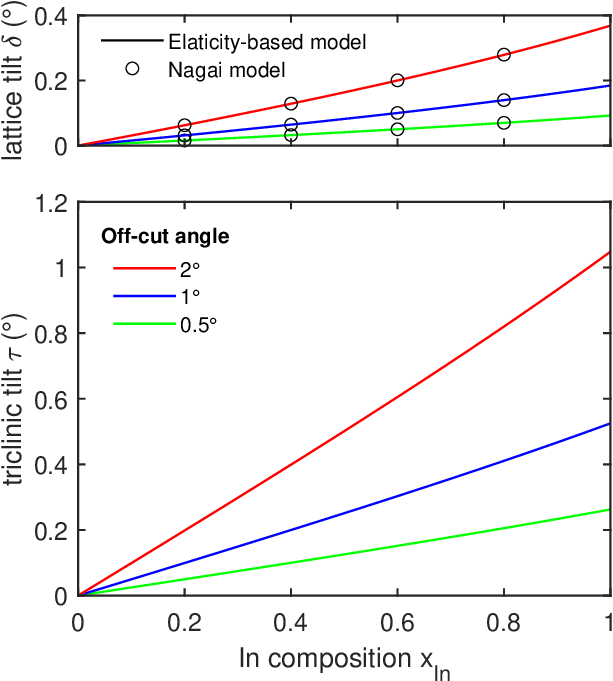}
\caption{Lattice tilt angle $\delta$ and triclinic tilt angle $\tau$ for a coherently strained In$_x$Ga$_{1-x}$N film on GaN(0001) as functions of composition $x_{\mathrm{In}}$ for off-cut angles of 0.5$^\circ$, 1$^\circ$, and 2$^\circ$. 
$\delta$ is calculated using both the Nagai and elasticity-based models, while $\tau$ is obtained from the elasticity-based model. 
The two models give nearly identical $\delta$, whereas $\tau$ reflects shear-induced deformation.
\label{fig:tilt_vs_off-cut_In}}
\end{figure}

As an example, we consider a coherently strained In$_x$Ga$_{1-x}$N film on a GaN substrate. Figure~\ref{fig:tilt_vs_off-cut_In} shows the calculated lattice tilt angle $\delta$ and triclinic tilt $\tau$ as functions of In composition $x_{\mathrm{In}}$ for off-cut angles of 0.5$^\circ$, 1$^\circ$, and 2$^\circ$. The calculations use elastic constants and lattice parameters of GaN and InN at $T=1076$~K\cite{ju2013situ,1997_Wright_JAP82_2833,2010_Richard_APL96_051911}, with interpolation for In$_x$Ga$_{1-x}$N based on Vegard's law.

As discussed in Appendix~\ref{sec:app.A}, the Nagai and elasticity-based models yield nearly identical values of $\delta$, indicating that the lattice tilt is primarily governed by normal strain, with shear effects entering only as higher-order corrections [Eq.~(\ref{eq:delta})]. In contrast, the elasticity-based model reveals an additional triclinic deformation, absent in the Nagai description. The magnitude of $\tau$ is found to be comparable to, or even larger than, $\delta$, demonstrating that shear-induced deformation is an essential component of the lattice distortion. As shown in Sec.~\ref{sb:IIIB}, this additional deformation has a pronounced effect on CTR scattering, particularly for non-specular rods, where sensitivity to in-plane lattice distortion is enhanced.

Having established the strained lattice geometry, we now incorporate it into the CTR scattering formalism through the epilayer structure factor and phase accumulation term.

\section{CTR scattering from vicinal surfaces under coherent heteroepitaxy}

In this section, we derive expressions for the X-ray reflectivity along CTRs from a vicinal substrate covered by a coherently strained heteroepitaxial film. Building on our previous formalism for vicinal surfaces without films\cite{2021_Ju_PRB_CTR}, we incorporate the complex scattering amplitude of the film with appropriate phase relations and lattice distortion. We then examine how the refined elastic crystal model modifies CTR profiles, particularly for non-specular rods that are sensitive to in-plane lattice distortion.

As in our previous work\cite{2021_Ju_PRB_CTR}, we employ the three-index orthogonal notation $H K L$, which provides a one-to-one mapping to the four-index hexagonal Miller-Bravais indices \(h k i \ell\) via \(H=h\), \(K=h+2k\), and \(L=\ell\).  All reciprocal-space coordinates are defined with respect to the substrate lattice. In Cartesian coordinates, the scattering vector $\mathbf{q}$= $(q_x,q_y,q_z)$ is given by $q_x = (2 \pi/a) H$, $q_y = (2 \pi/b) K$, $q_z = (2 \pi/c) L$. 
Substrate Bragg reflections occur at integer reciprocal lattice points $H_0 K_0 L_0$. 

The CTR intensity is calculated within the kinematic approximation, where multiple scattering is neglected, coupling  between the incident and scattered waves is ignored, and absorption is assumed to be weak. Within the Thomson scattering formalism, the complex reflectivity amplitude is given by\cite{1986_Robinson_PRB33_3830}
\begin{equation}
    r=\frac{i4\pi r_e}{AQ}\sum_j f_j(\mathbf{q})\exp(i\mathbf{q}\cdot\mathbf{r}_j),
\end{equation}
where $r_e = 2.817 \times 10^{-13}$~cm is the classical electron radius, $A$ is the in-plane unit-cell area, and $Q=|\mathbf{q}|$ is the magnitude of the scattering vector. $f_j(\mathbf{q})$ is the atomic form factor. The sum runs over every atom $j$. 

Considering lattice periodicity, the structure factor is defined as 
\begin{equation}
F(\mathbf{q})\equiv\sum_{\text{basis}}f_j(\mathbf{q})\exp(i\mathbf{q}\cdot\mathbf{r}_j^\prime),
\end{equation}
where $\mathbf{r}_j^\prime$ is the position of atom $j$ within the unit cell. Since $\mathbf{r}_j=\mathbf{R}+\mathbf{r}_j^\prime$, the total reflectivity can then be written as
\begin{equation}
r = r_\mathrm{f} F(\mathbf{q}) \sum_{\mathbf{R}} \exp(i\mathbf{q}\cdot\mathbf{R}),
\label{eq:gen}
\end{equation}
 where $r_\mathrm{f}=i4\pi r_e/(AQ)$. $\mathbf{R}$ runs over all lattice vectors of the crystal. The CTR calculation therefore reduces to evaluating the structure factor and the lattice phase sum. This formulation forms the basis for the CTR calculation presented below.

\subsection{\label{sb:IIIA}General CTR scattering formalism for vicinal surfaces }

\begin{figure}[t]
\includegraphics[width=1\linewidth]{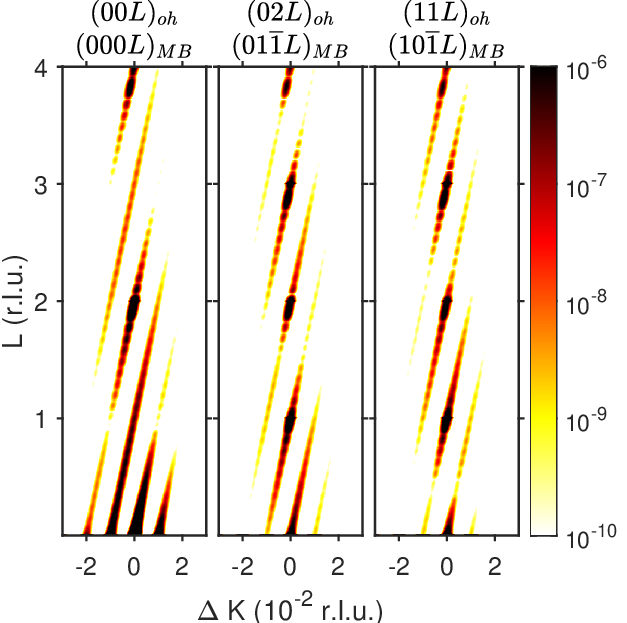}
\caption{Typical CTRs from a vicinal surface with a heteroepitaxial film, illustrated here using the InGaN/GaN system. The film introduces additional Bragg peaks and thickness fringes along the CTRs. \label{fig:KL}}
\end{figure}

For a vicinal surface with a coherent epitaxial film, the CTRs associated with Bragg peaks sharing the same in-plane indices $H_0K_0$ but different $L_0$ are directed along the surface normal. As a result, they are tilted with respect to the crystallographic axes by the off-cut angle $\theta$, as illustrated in Fig.~\ref{fig:KL}. Compared with an exactly oriented surface, the vicinal geometry separates rods that would otherwise overlap and provides enhanced sensitivity to step-resolved surface structure. Relative to our previous treatment of a vicinal substrate without an epitaxial layer\cite{2021_Ju_PRB_CTR}, the key new feature here is the emergence of interference fringes around each Bragg peak, arising from the finite thickness of the coherent epitaxial film. 

We consider the crystallographic model mentioned in Sec.~\ref{sec:II}, with steps occurring every $M$ unit cells along the $y$ direction. The total amplitude is decomposed into a semi-infinite substrate term and a finite-thickness epilayer term, shown by blue and red unit cells in Fig.~\ref{fig:InGaNonGaN}, both evaluated along the vicinal CTR defined by the step periodicity. The additional periodicity introduced by the step array leads to the characteristic tilted CTR pattern of a vicinal surface.

For the substrate, which is sufficiently thick to be treated as semi-infinite, the summation over bulk atoms [Eq.~(\ref{eq:gen})] is modulated by the additional periodicity introduced by the steps\cite{2002_Trainor_JApplCryst35_696}, and can be written as
\begin{align}  r_\mathrm{bulk}&=r_\mathrm{f}F_\mathrm{bulk}\sum_{\mathbf{R}} 
    \exp(i\mathbf{q}\cdot\mathbf{R})\nonumber\\
    &=r_\mathrm{f}F_\mathrm{bulk}\sum_{m=-\infty}^{0} \exp(iq_y mb)\nonumber\\
    \times&\sum_{N_3=-\infty}^{+\infty} \exp(i\mathbf{q} \cdot N_3(-M\mathbf{b} + \mathbf{c}))\sum_{N_1=-\infty}^{+\infty} \exp(iq_x N_1 a),
    \label{sum_step}
\end{align}
where the lattice vector is $\mathbf{R}=N_1 \mathbf{a}+m\mathbf{b}+N_3(-M\mathbf{b} + \mathbf{c})$. The index $m$ labels rows of unit cells parallel to the step edges, separated by one lattice spacing $b$ along the step-down direction. As illustrated by the filled blue unit cells in Fig.~\ref{fig:InGaNonGaN}(b), rows at the step edge correspond to $m=0$, while those displaced by one unit cell along the $-y$ direction correspond to $m=-1$, and so on. The last two summations in Eq.~(\ref{sum_step}) enforce phase coherence along the $x$ direction (via $N_1$) and the step periodicity (via $N_3$ along $-M\mathbf{b}+\mathbf{c}$), thereby confining the scattering to CTRs oriented along the vicinal surface normal, passing through successive Bragg points, as shown in Fig.~\ref{fig:KL}. 

The bulk structure factor, including thermal vibrations, is given by
\begin{align}
        F_\mathrm{bulk} &= \sum_{k} g_k(Q) \cdot \sum_{n_k}\exp(i\mathbf{q}\cdot\mathbf{r}_{k,n_k}^{\prime \text{ bulk}}),
    \label{eq:F_bulk}
\end{align}
where $g_k(Q) \equiv f_k(Q) \exp(-u_k^2 Q^2 / 2)$ is atomic form factor modified by Debye-Waller factor, with thermal vibration length $u_k$. Isotropic X-ray scattering factors $f_k(Q)$ are used so $\mathbf{q}$ is substituted by scalar $Q$. Within a unit cell, index $k$ labels atomic species and $n_k$ labels atoms within each species, thus each atom is identified by the pair $(k,n_k)$. $\mathbf{r}_{k,n_k}^{\prime \text{ bulk}}$ is the basis vector of atom $(k,n_k)$ in the bulk unit cell. 

To ensure convergence of the infinite summation in Eq.~(\ref{sum_step}), absorption is included. The bulk reflectivity amplitude becomes
\begin{align}
    r_\mathrm{bulk} &= \frac{r_\mathrm{f} \, F_\mathrm{bulk}}{M} \sum_{m = -\infty}^0 \exp (iq_y mb) \, [\exp (\epsilon \, b \tan \theta / q_z)]^m \nonumber\\
    &= \frac{r_\mathrm{f} \, F_\mathrm{bulk}}{M} \frac{Y_\mathrm{bulk}}{Y_\mathrm{bulk} - 1}, 
    \label{eq:r_b_v} 
\end{align}
where
\begin{align}
    Y_\mathrm{bulk} & \equiv \exp (iq_y b) \, \exp (\epsilon \, b \tan \theta / q_z) \nonumber \\
    & = \exp (2 \pi i K) \, \exp (\epsilon c^2 / 2 \pi M L). 
\end{align}
Here, $\epsilon = 4\pi / (\lambda \ell_{\text{abs}})$, with $\lambda$ the X-ray wavelength and $\ell_{\text{abs}}$ the absorption length. Although the exponential absorption factor is typically close to unity, it mathematically regularizes the divergence of the bulk summation at Bragg peaks. The factor $1/M$ in Eq.~(\ref{eq:r_b_v}) arises from the summation over the step periodicity ($N_3$) in Eq.~(\ref{sum_step}) and ensures proper normalization of the scattering amplitude. For a given Bragg peak $H_0 K_0 L_0$, the corresponding CTR satisfies 
\begin{equation}
K = K_0 + \frac{L - L_0}{M},
\label{eq:K}
\end{equation}
reflecting its orientation along the vicinal surface normal. In the limit $M\to \infty$, corresponding to an exactly oriented surface, the expression recovers the conventional CTR result\cite{2021_Ju_PRB_CTR}. Substituting this relation into the phase factor eliminates the explicit dependence on  $K_0$, yielding
\begin{equation}
    Y_{\mathrm{bulk}} = \exp [2 \pi i (L - L_0) / M] \exp (\epsilon c^2 / 2 \pi M L).
\end{equation}

We now extend this formalism to include the contribution from a coherent epitaxial film. The scattering amplitude is obtained by continuing the summation in Eq.~(\ref{eq:r_b_v}) from the substrate region ($m \leq 0$) into the film region ($m=1$ to $JM$), thereby preserving the phase continuity across the substrate-film interface while introducing distinct phase factors in the epilayer containing lattice distortion.  This yields
\begin{equation}
    r_\mathrm{epi} = \frac{r_\mathrm{f} \, F_\mathrm{epi}}{M} \sum_{m = 1}^{JM} Y_\mathrm{epi}^m = \frac{r_\mathrm{f} \, F_\mathrm{epi}}{M} \frac{Y_\mathrm{epi}(Y_\mathrm{epi}^{JM}-1)}{Y_\mathrm{epi} - 1}, 
    \label{eq:r_epi_v} 
\end{equation}
where $J$ denotes the film thickness in unit cells, so that $JM$ gives the total number of unit cells along $y$ in the epilayer. For a vicinal surface, $J$ need not be integer; only $JM$ must be integer.

The epilayer structure factor, including elastic strain and lattice rotation, is
\begin{equation}
\begin{aligned}
F_\mathrm{epi} &= \sum_{k} g_k(Q) \sum_{n_k}
\exp\!\left(i\mathbf{q}\cdot\boldsymbol{\Delta}'\cdot\mathbf{r}_{k,n_k}^{\prime 0}\right), \\
\text{with}\quad
\boldsymbol{\Delta}' &= \boldsymbol{T}_x(\gamma)(\boldsymbol{I}+\boldsymbol{\epsilon}),
\end{aligned}
\label{eq:F_epi_2}
\end{equation}
where $\boldsymbol{\Delta}'$ describes the full lattice distortion of the epilayer, including both elastic strain and rigid-body rotation (lattice tilt). Here $\mathbf{r}_{k,n_k}^{\prime\text{0}}$ denotes the atomic basis positions in the unstrained epilayer, so that $\boldsymbol{\Delta}'$ acts directly as the distortion tensor in Cartesian coordinates. The corresponding phase factor becomes 
\begin{equation}
    Y_\mathrm{epi}  \equiv \exp (i \mathbf{q}\cdot\mathbf{b}_\mathrm{epi}) \, \exp (\epsilon_\mathrm{epi} \, b_\mathrm{epi} \tan \theta_\mathrm{epi}  / q_z),
    \label{eq:Yes}
\end{equation}
where $\mathbf{b}_\mathrm{epi}$ is no longer parallel to the bulk lattice vector  $\mathbf{b}$. For the finite film thickness considered here, variations in the absorption term are negligible.

The total reflectivity amplitude is the sum of the complex amplitudes from the bulk and the film:
\begin{equation}
r_\mathrm{tot}=r_\mathrm{bulk}+r_\mathrm{epi}.
\end{equation}
 
To obtain the measurable reflectivity, we further include both dynamical corrections near the Bragg condition and surface roughness effects. Near the Bragg peaks, where the reflectivity amplitude approaches unity, the kinematic approximation breaks down. The complex amplitude is therefore corrected using the dynamical expression\cite{1997_Thompson_APL71_3516}
\begin{equation}
    r_\mathrm{tot}^{dyn} = \frac{2 r_\mathrm{tot}}{1 + \sqrt{1 + 4 |r_\mathrm{tot}|^2}},
    \label{eq:dyn}
\end{equation}
which refines the peak profile and enforces the physically required bound on the reflectivity, preventing the unphysical divergence inherent to the kinematic approximation. 

Surface roughness leads to an attenuation of the CTR intensity away from the Bragg condition. The reflectivity is therefore written as
\begin{equation}
R(L) = |r_\mathrm{tot}^{dyn}(L)|^2 \, S_{L_0}(L),
\label{eq:R}
\end{equation}
where the roughness factor is described by a Gaussian form\cite{1985_Andrews_JPhysC18_6427,1997_Munkholm_JApplPhys82_2944,2017_Petach_PRB95_184104},
\begin{equation}
    S_{L_0} = \exp \big [ - \sigma_R^2 (2\pi/c)^2 (L - L_0)^2 \big ],
    \label{eq:SL0}
\end{equation}
with $\sigma_R$ denoting the root-mean-square surface roughness. Here, each CTR is analyzed independently in the vicinity of a given $L_0$. This treatment is justified in the small off-cut limit (large $M$) of the vicinal surface, where contributions from neighboring Bragg peaks are well separated in reciprocal space, as illustrated in Fig.~\ref{fig:KL}.

Finally, real surfaces often exhibit additional structural complexities, such as reconstruction, terrace ordering, and local compositional variations, all of which can significantly modify the CTR intensity. Such effects can be incorporated into the present formalism through additional surface-specific contributions to $r_\mathrm{tot}$.

In Sec.~\ref{sec:IV}, we apply this framework to the InGaN/GaN system, explicitly accounting for terrace-resolved reconstruction, $\alpha/\beta$ terrace ordering, and step-dependent indium incorporation, and demonstrate how these factors quantitatively modify CTR profiles.

\subsection{\label{sb:IIIB}Effect of triclinic deformation on CTR scattering}

\begin{figure*}
\centering
\includegraphics[width=0.85\linewidth]{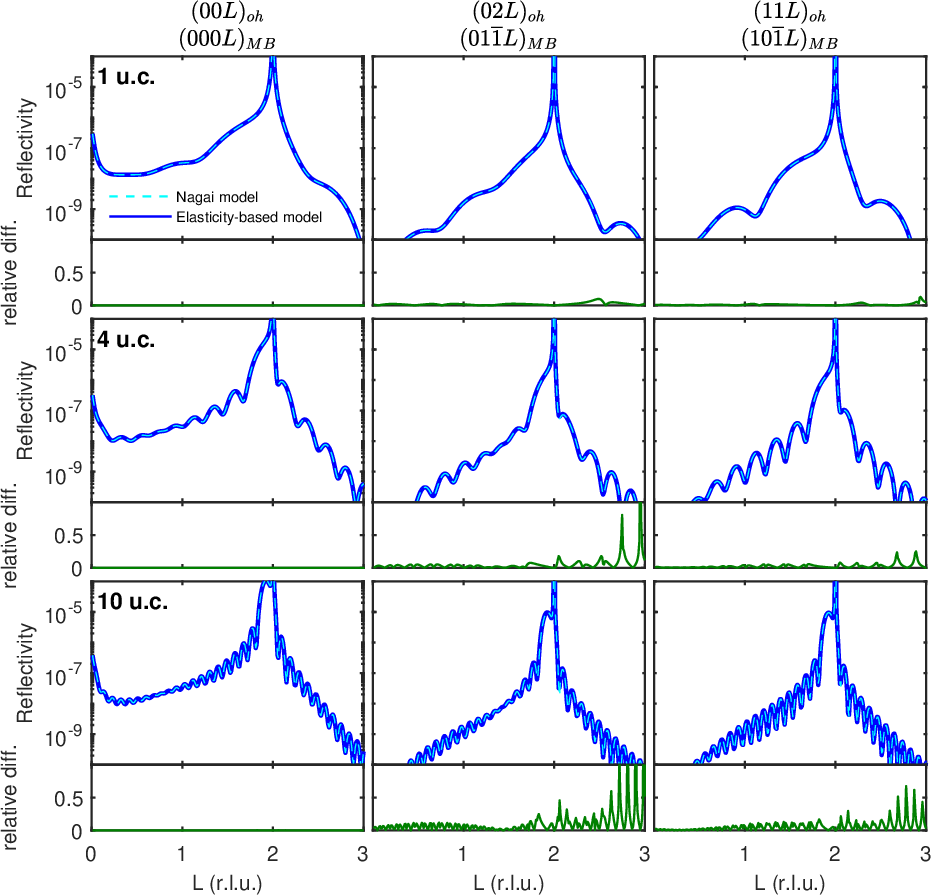} 
\caption{Calculated reflectivities for the $(0 0 L_0)$, $(0 2 L_0)$, and $(1 1 L_0)$ CTRs (from left to right) at $L_0=2$ for a vicinal GaN(0001) surface with $M=100$, covered by a coherent In$_{0.3}$Ga$_{0.7}$N film. Rows correspond to film thicknesses of 1, 4, and 10 unit cells (top to bottom). The lower panels show the relative difference between the Nagai and elasticity-based model. The non-specular CTRs are markedly more sensitive to triclinic deformation than the specular CTR, and the this sensitivity increases with film thickness. \label{fig:CTR_mdl_dif} }
\end{figure*}

The comparison between the Nagai and elasticity-based model in Sec.~\ref{sec:II} shows that both models yield nearly identical lattice tilt angles $\delta$, but differ in the presence of an additional triclinic deformation characterized by $\tau$. This distinction has little effect on specular CTRs, but becomes significant for non-specular rods, where the scattering is sensitive to in-plane lattice distortion.

The origin of this behavior lies in the epilayer contribution $r_\mathrm{epi}$ in Eq.~(\ref{eq:r_epi_v}), since the bulk contribution is identical in both models. The difference arises from the phase term $\mathbf{q}\cdot\boldsymbol{\Delta}'$, entering both the structure factor $F_\mathrm{epi}$ [Eq.~(\ref{eq:F_epi_2})] and the interference term $Y_\mathrm{epi}$ [Eq.~(\ref{eq:Yes})].

In the Nagai model, the deformation matrix is given by 
$\boldsymbol{\Delta'}_N = \boldsymbol{T}_x(\delta)\,\mathrm{diag}
\left(
a_\mathrm{epi}/a_\mathrm{epi}^0,\,
b_\mathrm{epi}/b_\mathrm{epi}^0,\,
c_\mathrm{epi}/c_\mathrm{epi}^0
\right)$,
which includes only lattice tilt and normal strain. In contrast, the elasticity-based model,
$\boldsymbol{\Delta'}_e = \boldsymbol{T}_x(\gamma)(\boldsymbol{I}+\boldsymbol{\epsilon})$,
incorporates the full strain tensor and therefore includes off-diagonal components associated with in-plane shear. 
To explicitly quantify this difference, we evaluate the matrix difference $\boldsymbol{\Delta'}_e - \boldsymbol{\Delta'}_N$ (see Appendix~\ref{app:Small-quantity}).

For representative parameters ($\mathrm{In_{0.3}Ga_{0.7}N}$ at 1076K with $M=100$, corresponding $\theta=0.54^\circ$), the difference between the two models is explicitly quantified by evaluating the dimensionless difference matrix 
\begin{equation}
\boldsymbol{\Delta^\prime_e}-\boldsymbol{\Delta^\prime_N}=
\left[
\begin{array}{ccc}
0 & 0 & 0\\
0 & 1\times10^{-5} & 0.0014\\
0 & -5\times10^{-8} & -6\times10^{-6}
\end{array}
\right],
\end{equation}
which shows that the dominant contribution is the (2,3) component, while all other terms are negligible.

The difference of $F_\mathrm{epi}$ is governed by the the distorted-lattice phase term $\mathbf{q}\cdot\boldsymbol{\Delta'}$, where $\mathbf{q}=[2\pi H/a,\;2\pi K/b,\;2\pi L/c]$. For the specular $(00L)$ rod, $q_x = q_y = 0$, so the phase factors are insensitive to in-plane shear, suppressing its contribution and resulting in nearly identical reflectivities for the two models.

In contrast, for non-specular rods ($K\neq 0$), the in-plane component $q_y$ directly couples to the shear terms in $\boldsymbol{\Delta'}$, producing a significant phase shift and thereby a pronounced difference in the CTR profiles.

The effect is further amplified by the finite film thickness. The interference factor
\begin{equation}
Y_\mathrm{epi}=\exp (i \mathbf{q}\cdot\boldsymbol{\Delta}'\cdot\mathbf{b}_\mathrm{epi}^0) \, 
\exp (\epsilon_\mathrm{epi} \, b_\mathrm{epi} \tan \theta_\mathrm{epi} / q_z),
\end{equation}
since $\mathbf{b}_\mathrm{epi}=\boldsymbol{\Delta}'\cdot\mathbf{b}_\mathrm{epi}^0$. For the same representative case,
\begin{align}
\mathbf{b}_{\mathrm{epi},e}-\mathbf{b}_{\mathrm{epi},N}
&=(\boldsymbol{\Delta'}_e - \boldsymbol{\Delta'}_N)\cdot\mathbf{b}_\mathrm{epi}^0\nonumber\\
&=[0,\,7.6\times10^{-5},\,-3.2\times10^{-7}]^T,
\end{align}
which shows that the dominant difference is in the in-plane component, leading to a cumulative phase difference through the interference term $(Y_\mathrm{epi}^{JM}-1)/(Y_\mathrm{epi}-1)$.

Figure~\ref{fig:CTR_mdl_dif} illustrates these trends. The specular $(00L)$ CTR shows no discernible difference between the two models, whereas the non-specular rods exhibit clear deviations in both peak shape and interference fringes. The discrepancy grows with film thickness because the shear-induced phase shift accumulates through the finite-thickness interference factor.

These results show that non-specular CTRs provide a sensitive probe of triclinic lattice deformation (in-plane shear) in coherent heteroepitaxial films, while such effects remain essentially invisible in specular measurements.

\section{\label{sec:IV}Step-resolved CTR signatures of coherent heteroepitaxy on vicinal surfaces}

\begin{figure}[b]
\includegraphics[width=1\linewidth]{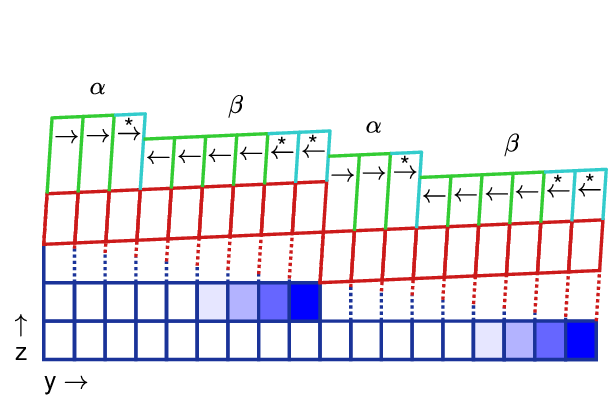}
\caption{Schematic illustration of the unit-cell configuration on a vicinal surface. Bulk unit cells (blue), internal film unit cells (red), and top film unit cells with reconstruction (green) or modified reconstruction (cyan) are shown. The total terrace width is composed of $\alpha$ and $\beta$ terraces, with $M = 9$ unit cells. The $\alpha$ terrace has width $N = 3$, within which a modified region of width $N_{\alpha^*} = 1$ is present. The $\beta$ terrace contains a modified region of width $N_{\beta^*} = 2$. The film thickness is $J = 3$ unit cells. Arrows indicate the reconstruction orientation, which alternates between $\alpha$- and $\beta$-terminated terraces. The blue shading highlights the summation index $m$ defined in Eq.~(\ref{rec_epi}), with the darkest region corresponding to $m = 0$.\label{fig:recon}}
\end{figure}

In this section, we extend the formalism to resolve terrace-specific contributions to the CTR intensity, including $\alpha/\beta$ terrace ordering and surface reconstruction.

\begin{figure*}
\centering
\includegraphics[width=0.9\linewidth]{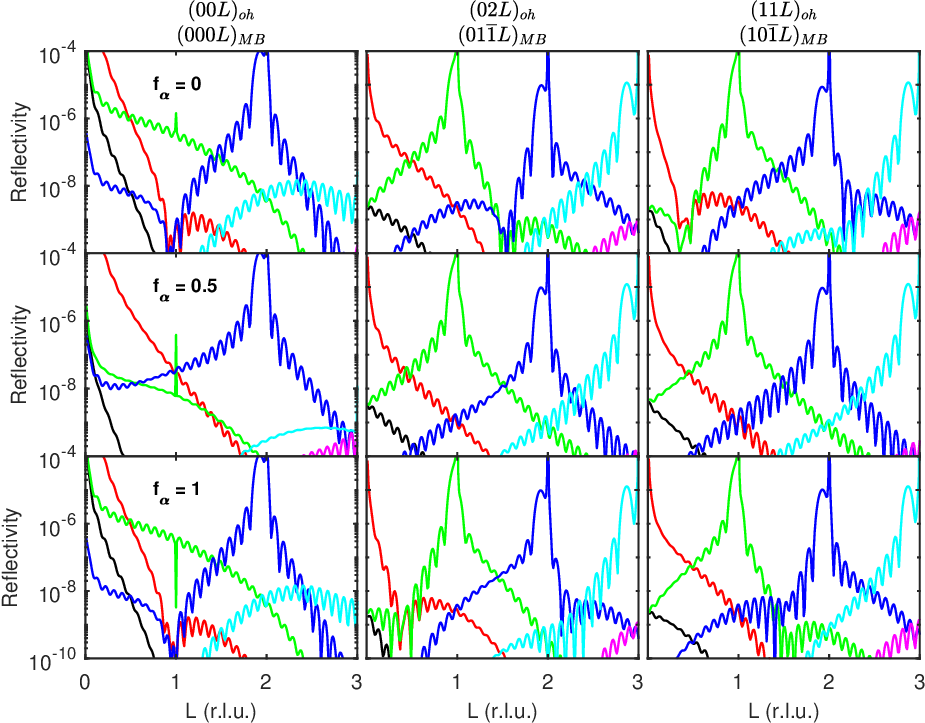}
\caption{Calculated reflectivities for the $(00L_0)$, $(02L_0)$, and $(11L_0)$ CTRs for a vicinal GaN(0001) surface with $M=100$, covered by a coherent In$_{0.3}$Ga$_{0.7}$N epitaxial film with a thickness of 10 unit cells. Colors denote different CTRs with $L_0=-1$ to $4$, and rows correspond to different $\alpha$-terrace fractions $f_\alpha$. \label{fig:CTR_off-cut} }
\end{figure*}

\subsection{CTR sensitivity to surface reconstruction and terrace ordering}

\begin{figure*}
\centering
\includegraphics[width=1\linewidth]{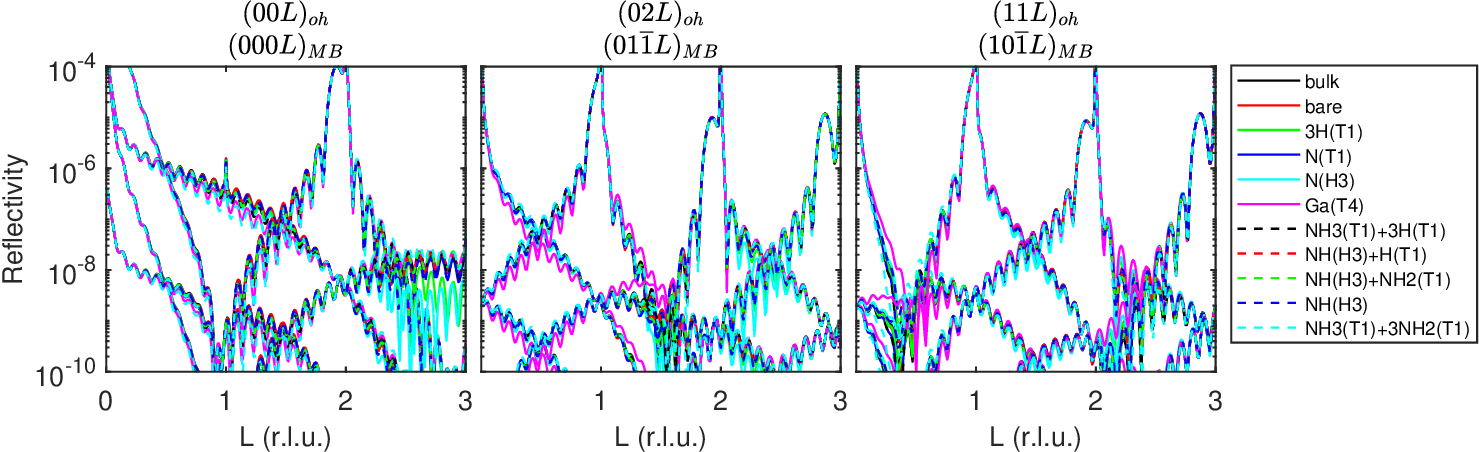}
\caption{Calculated CTR reflectivities for a vicinal GaN(0001) surface with $M = 100$ and a 10-unit-cell thick coherent In$_{0.3}$Ga$_{0.7}$N epitaxial film, for various reconstructions\cite{inatomi2020theoretical} at $f_\alpha = 0$. 
\label{fig:CTR_miscut_recon} }
\end{figure*}

Surface reconstruction modifies the CTR intensity through changes in the surface structure factor. Owing to its sensitivity to atomic-scale surface structure, CTR scattering has been widely used to determine surface reconstructions in semiconductor\cite{1990_van_FaradayDiscuss,2021_Ju_NatCommun12_1721} and oxide systems\cite{1998_Guenard_SurfRevLett,2000_Eng_Science}. 

In addition to reconstruction, vicinal surfaces introduce further complexity through $\alpha$/$\beta$ terrace ordering in the hexagonal system \cite{1999_Xie_PRL82_2749, 2021_Ju_PRB_CTR}, which can also strongly affect the CTR profiles. To isolate and quantify these effects, we extend the general model by incorporating reconstructed surface layers and terrace-resolved contributions.

In this framework, we adopt a surface-resolved description of the scattering amplitude, in which the topmost unit-cell layer is treated explicitly as part of the reconstructed surface. The bulk contribution $r_\mathrm{bulk}$ remains unchanged from Sec.~\ref{sb:IIIA}, since it is independent of the surface structure. The total reflectivity amplitude is then expressed as a sum of bulk, film, and surface contributions.

For a uniform surface reconstruction without terrace ordering, the total reflectivity amplitude is given by
\begin{equation}
r_\mathrm{tot}=r_\mathrm{bulk}+r_\mathrm{epi}+r_\mathrm{rec},
\end{equation}
where $r_\mathrm{rec}$ represents the contribution from a uniformly reconstructed surface layer.

For surfaces with alternating $\alpha$ and $\beta$ terraces, the surface contribution is further decomposed into terrace-resolved components, and the total reflectivity becomes
\begin{equation}
    r_\mathrm{tot}=r_\mathrm{bulk}+r_\mathrm{epi}+f_{\alpha}r_{\alpha}+(1-f_{\alpha})r_\beta,
    \label{eq:rtot_ab}
\end{equation}
where $r_\alpha$ and $r_\beta$ denote the contributions from the corresponding terrace types, and $f_\alpha$ is the fractional surface coverage of the $\alpha$ terraces. As illustrated in Fig.~\ref{fig:recon}, $f_\alpha$ is given by $N/M$, where $N$ and $M-N$ represent the widths of the $\alpha$ and $\beta$ terraces, respectively, within one surface period $M$. 

In this surface-resolved formulation, the topmost unit-cell layer is assigned to the reconstructed surface and is therefore excluded from the epilayer contribution. Accordingly, the epilayer term includes only the underlying $(J-1)$ unit cells, and can be written as
\begin{equation}
    r_\mathrm{epi}
    =
    \frac{r_\mathrm{f}F_\mathrm{epi}}{M}
    \sum_{m=1}^{(J-1)M}Y_\mathrm{epi}^m
    =
    \frac{r_\mathrm{f}F_\mathrm{epi}}{M}
    \frac{Y_\mathrm{epi}\left(1-Y_\mathrm{epi}^{(J-1)M}\right)}{1-Y_\mathrm{epi}}.
    \label{rec_epi}
\end{equation}
This redefinition leaves the total reflectivity unchanged and only redistributes the amplitude between the epilayer and surface terms.

For a uniform surface reconstruction, the surface contribution takes the form 
\begin{align}
    r_{\mathrm{rec}}&= \frac{r_\mathrm{f} \, F_{\mathrm{rec,u}}}{M} \sum_{m = (J-1)M+1}^{(J-1)M+M} Y_\mathrm{rec}^m \nonumber \\
    &= \frac{r_\mathrm{f} \, F_\mathrm{rec,u}}{M} \frac{Y_\mathrm{rec}^{(J-1)M+1}(1-Y_\mathrm{rec}^{M} )}{1-Y_\mathrm{rec}},
    \label{eq:r_rec}
\end{align}
where $Y_\mathrm{rec}$ can be taken equal to $Y_{\mathrm{epi}}$[Eq.~(\ref{eq:Yes})], since the absorption term is close to unity for a surface-localized contribution, and the phase term $\exp(i\mathbf{q}\cdot\mathbf{b}_\mathrm{epi})$ remains the same.

In general, multiple surface reconstructions and domain orientations may coexist. Assuming that each domain is small compared with the X-ray beam footprint, the total surface contribution is obtained by averaging over all domains,
\begin{equation}
    r_\mathrm{rec} = \sum_{\phi,\xi} f_{\phi \xi} \, r_{\mathrm{rec},\phi \xi},
\end{equation}
where $\phi$ labels symmetry-equivalent orientations, $\xi$ labels reconstruction types, and $f_{\phi \xi}$ is the corresponding surface fraction. The coefficients $f_{\phi \xi}$ can be absorbed into an effective structure factor
\begin{equation}
    F_\mathrm{rec}=\sum_{\phi,\xi} f_{\phi \xi} \, F_{\mathrm{rec},\phi \xi},
\end{equation}
so that $r_\mathrm{rec}$ retains the same form as Eq.~(\ref{eq:r_rec}).

For the surface with alternating $\alpha$ and $\beta$ terraces, the reflectivity contribution from each terrace type ($w=\alpha,\beta$) can be written as
\begin{align}
    r_w &= \frac{r_\mathrm{f} \, F_w}{M_w} \sum_{m = M_{w0}+1}^{M_{w0}+M_w} Y_\mathrm{epi}^m \nonumber \\
    &= \frac{r_\mathrm{f} \, F_w}{M_w} \frac{Y_\mathrm{epi}^{M_{w0}+1}(1-Y_\mathrm{epi}^{M_w} )}{1-Y_\mathrm{epi} },
    \label{eq:rx}
\end{align}
where $M_{\alpha 0}=(J-1)M$, $M_{\beta 0}=(J-1)M+N$, $M_\alpha=N$, and $M_\beta=M-N$.

The structure factor for each terrace type is
\begin{align}
    & F_w = \sum_\phi f_{\phi w} \sum_k g_k(Q) \sum_n \exp(i \mathbf{q} \cdot \boldsymbol{\Delta}' \cdot  \mathbf{r}_{\phi kn}^{w\prime}), \label{eq:F_recm}
\end{align}
where $k$ labels atomic species, and $n$ labels atoms of type $k$ within the unit cell. The unit cell differs for $\alpha$ and $\beta$ terraces due to their distinct stacking sequences and reconstruction configurations, as shown in Fig.~\ref{fig:recon}. 

As a representative example, we consider CTR calculations for a vicinal GaN(0001) surface with a coherent InGaN epitaxial film under experimentally relevant conditions. The out-of-plane lattice parameter of the film and its reconstructed surface, $c_\mathrm{epi}$, is determined assuming coherent strain, using lattice parameters and elastic constants reported in Ref.~\citenum{2010_Richard_APL96_051911}. For all calculations presented in this work, the growth temperature is fixed at $1076\,\mathrm{K}$. The surface reconstruction is taken to be 3H(T1)\cite{2021_Ju_NatCommun12_1721}. A photon energy of $25.78$~keV ($\lambda = 0.4809$~\AA) is used, consistent with recent experiments.\cite{2021_Ju_NatCommun12_1721} Atomic form factors are taken from Ref.~\citenum{1995_Waasmaier_ActaCrystA51_416}, including resonant corrections at this energy.\cite{1993_Henke_ANDT54_181} Absorption lengths of $\ell_\mathrm{abs} = 101$ and $172\,\mu\mathrm{m}$ are used for GaN and InN, respectively.\cite{CXRO} A Debye–Waller length of $u_k = 0.16$~\AA~ is assumed for all atoms, and the surface roughness is taken to be $\sigma_R = 1$~\AA.

Figure~\ref{fig:CTR_off-cut} shows the calculated CTR reflectivities for a vicinal GaN(0001) surface with a coherent InGaN film. Three families of CTRs are considered: $(0 0 L_0)$, $(0 2 L_0)$, and $(1 1 L_0)$, for $L_0 = -1$ to $4$, with varying terrace fractions $f_\alpha$. The overall behavior is similar to that for a GaN surface with no epitaxial film, modulated by the extra Bragg peaks and thickness fringes due to the film\cite{2021_Ju_PRB_CTR}.
The $(0 0 L_0)$ CTRs exhibit a symmetric dependence on $f_\alpha$ about $f_\alpha = 0.5$. The profiles for $f_\alpha = 0$ and $f_\alpha = 1$ are nearly identical, while a pronounced modulation appears at $f_\alpha = 0.5$, where the intensities at even $L_0$ are enhanced and those at odd $L_0$ are suppressed. 

In contrast, the $(0 2 L_0)$ and $(1 1 L_0)$ CTRs show a strong and monotonic dependence on $f_\alpha$. Focusing on the $(0 2 L_0)$ rod with $L_0=2$, a dip appears on the low-$L$ side of the Bragg peak for $f_\alpha = 0$. As $f_\alpha$ increases, this dip gradually shifts and reverses, leading to a mirrored feature on the high-$L$ side for $f_\alpha = 1$. The $(1 1 L_0)$ CTRs exhibit the opposite trend. This complementary behavior reflects the symmetry relation between the two rods: a 30$^\circ$ in-plane rotation exchanges the $(0 2 L)$ and $(1 1 L)$ directions while interchanging $\alpha$ and $\beta$ terraces.

\begin{figure*}
\centering
\includegraphics[width=0.9\linewidth]{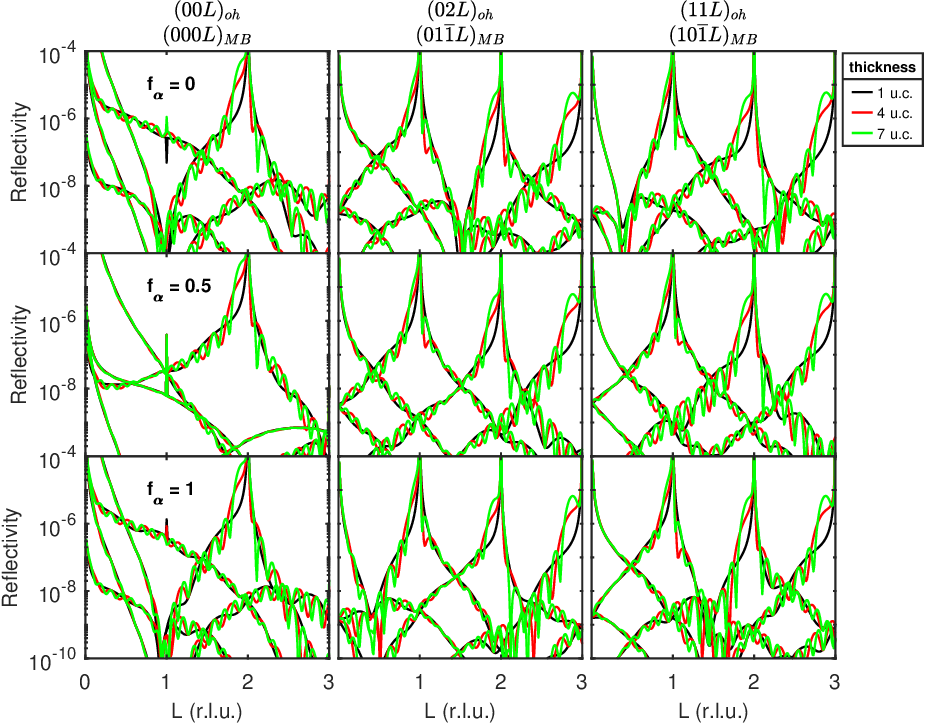}
\caption{Calculated reflectivities for the $(00L_0)$, $(02L_0)$, and $(11L_0)$ CTRs for $L_0=-1$ to $4$ of a vicinal GaN(0001) surface with $M=100$, covered by a coherent In$_{0.3}$Ga$_{0.7}$N epitaxial film. Rows correspond to $\alpha$-terrace fractions $f_\alpha$ (top to bottom). Colors correspond to film thicknesses of 1, 4, and 7 unit cells. 
\label{fig:CTR_off-cut2} }
\end{figure*}

Surfaces with different reconstructions exhibit the similar qualitative dependence on $f_\alpha$, as shown in Fig.~\ref{fig:CTR_miscut_recon}. This indicates that the terrace ordering effect is largely independent of the specific reconstruction details. For each reconstruction, the corresponding surface structure factor is obtained from DFT, and the CTR intensity is calculated accordingly. Comparison with experimental data through $\chi^2$ minimization allows identification of the most probable surface reconstruction\cite{2021_Ju_NatCommun12_1721}.
Furthermore, as illustrated in Fig.~\ref{fig:CTR_off-cut2}, the qualitative dependence on $f_\alpha$ is largely independent of film thickness, demonstrating that the observed behavior arises from the geometric arrangement of terraces rather than thickness-related interference effects.

\subsection{Terrace-resolved surface indium enrichment and step-selective incorporation}

\begin{figure*}
\includegraphics[width=0.9\linewidth]{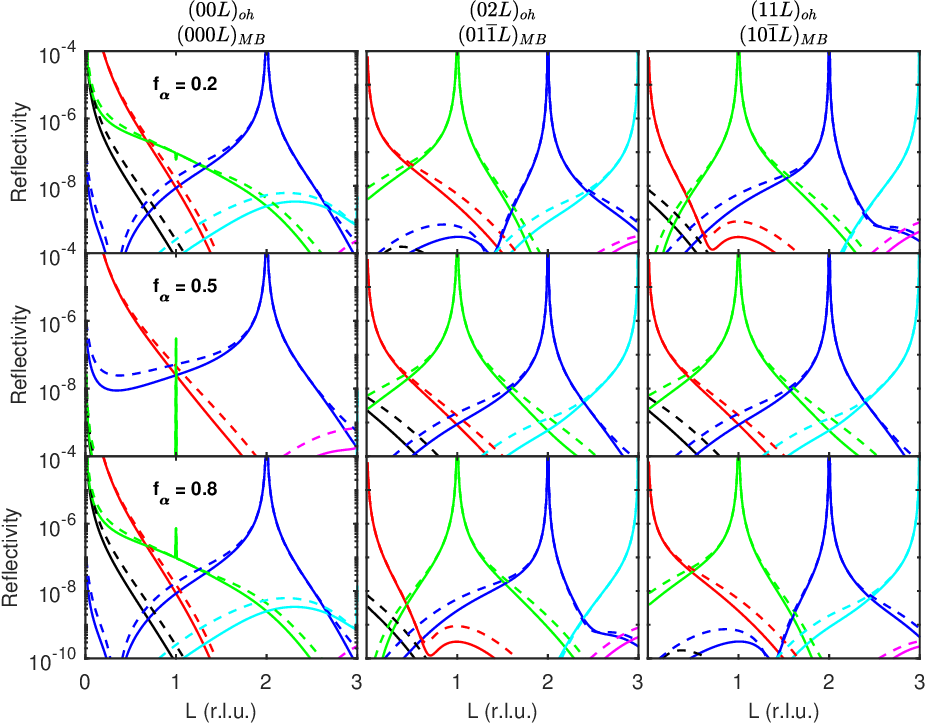}
\caption{Calculated reflectivities for the $(00L_0)$, $(02L_0)$, and $(11L_0)$ CTRs for a vicinal GaN(0001) surface without an epitaxial film, showing only the effect of a modified surface reconstruction with $x_{\mathrm{In},\alpha}^* = x_{\mathrm{In},\beta}^* = 1$. Solid and dashed curves correspond to $f_{\alpha^*} = f_{\beta^*} = 0$ and $0.2$, respectively. Colors denote different CTRs with $L_0=-1$ to $4$. Rows correspond to different $\alpha$-terrace fractions $f_\alpha$.
\label{fig:CTR_off-cut_ad1}}
\end{figure*}

Previous theoretical work has suggested that indium incorporation during III-nitride growth may proceed via preferential adsorption at specific step edges, followed by incorporation into the lattice, leading to lateral compositional non-uniformity in the growing film\cite{uvzdavinys2018impact}. However, experimental techniques capable of quantitatively resolving such step-selective incorporation processes remain lacking. Given its intrinsic sensitivity to atomic-scale surface structure and terrace-specific contributions, CTR scattering provides a promising route to probe this long-standing problem. We first analyze step-selective indium incorporation in the film-free limit within a theoretical framework that can be tested in future experiments.

Selective indium incorporation modifies the local surface structure and results in reconstructed regions with altered indium composition, schematically illustrated by the cyan lattice in Fig.~\ref{fig:recon}. Building upon the $\alpha/\beta$ terrace-resolved CTR formalism developed above, we introduce modified reconstruction domains that account for terrace-dependent indium incorporation. Specifically, the modified reconstructions on the $\alpha$ and $\beta$ terraces are denoted as $\alpha^*$ and $\beta^*$, respectively. Within a double-step spacing of $M$ unit cells in the $y$ direction, the modified regions occupy $N_{\alpha^*}$ and $N_{\beta^*}$ unit cells on the $\alpha$ and $\beta$ terraces, respectively. The corresponding surface fractions are defined as $f_{\alpha^*} = N_{\alpha^*}/M$ and $f_{\beta^*} = N_{\beta^*}/M$, representing the fractions of the surface occupied by the modified reconstructions on the respective terraces. 

The total reflectivity amplitude is then expressed as the coherent sum of contributions from the bulk, the epilayer, and the terrace-resolved reconstructed regions:
\begin{align}
    r_\mathrm{tot} &= r_\mathrm{bulk} + r_\mathrm{epi} \nonumber \\
    &+ (f_\alpha - f_{\alpha^*}) \, r_\alpha + f_{\alpha^*} \, r_{\alpha^*} \nonumber \\
    &+ (1 - f_\alpha - f_{\beta^*}) \, r_\beta + f_{\beta^*} \, r_{\beta^*},
    \label{eq:rt}
\end{align}
where $f_\alpha^* \leq f_\alpha$ and $f_\beta^* \leq 1 - f_\alpha$.

The reflectivity amplitudes of the individual terrace regions ($w = \alpha, \alpha^*, \beta, \beta^*$) are calculated using Eq.~(\ref{eq:rx}), where the summation is performed along the step-down ($y$) direction over the unit cells occupied by each reconstruction. The corresponding summation ranges are defined by the start index $M_{w0}$ and length $M_w$ for each region, as listed in Table~\ref{tab:mx}. These definitions ensure that the terrace-resolved contributions are coherently summed within a single periodic repeat unit of length $M$, preserving the phase relationships required for CTR interference. The structure factor for each reconstruction retains the general form given in Eq.~(\ref{eq:F_recm}), with modified basis atomic positions $\mathbf{r}_{\phi kn}^{\prime w}$ used to account for the indium-enriched configurations in the $\alpha^*$ and $\beta^*$ domains.

\begin{table}[b] 
\caption{ \label{tab:mx} Values of start and length coefficients $M_{w0}$ and $M_w$ in Eq.~(\ref{eq:rx}) for each reconstruction $w = \alpha$, $\alpha^*$, $\beta$, and $\beta^*$. $M_{w0}$ gives the starting row index of region $w$; $M_w$ gives the number of rows in that region. }
\begin{ruledtabular}
\begin{tabular}{ c  c c} 
$w$ & $M_{w0}$ & $M_w$ \\ \hline
$\alpha$ & $(J-1)M$  & $N - N_{\alpha^*}$   \\
$\alpha^*$ & $(J-1)M + N - N_{\alpha^*}$  & $N_{\alpha^*}$   \\
$\beta$ & $(J-1)M + N$  & $M - N - N_{\beta^*}$   \\
$\beta^*$ & $(J-1)M + M - N_{\beta^*}$  & $N_{\beta^*}$ 
\end{tabular}
\end{ruledtabular}
\end{table}

To clarify the intrinsic CTR response to terrace-resolved surface indium enrichment, we first consider the simplest case without an epitaxial film. This avoids the additional thickness fringes introduced by film scattering and isolates the effect of modified surface reconstruction near the step edges. Figure~\ref{fig:CTR_off-cut_ad1} shows the calculated CTR profiles for a vicinal GaN(0001) surface in this limit. In this example, the modified regions correspond to indium-rich top half unit cells ($x_{In,\alpha}^* = x_{In,\beta}^* = 1$) with a total coverage of 40\% ($f_{\alpha^*} = f_{\beta^*} = 0.2$) distributed over a pair of adjacent steps. Compared with the unmodified case, the incorporation of indium leads to a pronounced increase in CTR intensity away from the Bragg peaks. This effect is consistently observed across different CTR types and $\alpha$ terrace fractions $f_{\alpha}$.

These results show that CTR scattering is sensitive to terrace-resolved surface modifications associated with selective indium incorporation. The calculations suggest that, in principle, simultaneous fitting of the terrace fraction and the coverage of indium-enriched surface regions should be feasible, provided that data of sufficient quality are available. In practical applications, this sensitivity offers a route to quantitatively probe step-dependent incorporation processes and local surface composition, including terrace-dependent indium enrichment and its relation to segregation phenomena.

\begin{figure*}
\includegraphics[width=0.9\linewidth]{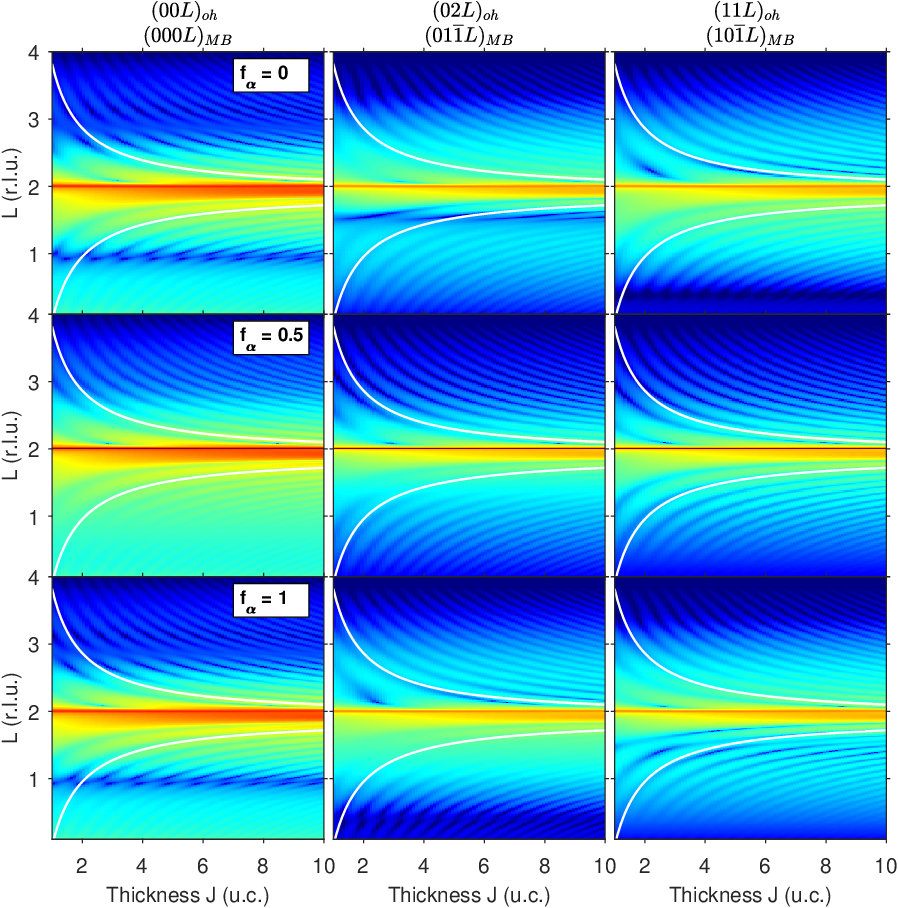}
\caption{Color map of the logarithm of the CTR reflectivity for the $(00L_0)$, $(02L_0)$, and $(11L_0)$ CTRs at $L_0=2$, for a vicinal GaN(0001) surface with a coherent In$_{0.3}$Ga$_{0.7}$N epitaxial film. The intensity is shown as a function of film thickness and $L$ in the vicinity of the $(H_0 K_0 L_0)$ Bragg peak. Rows correspond to different $\alpha$-terrace fractions $f_\alpha$. The white curves indicate equiphase trajectories given by Eq.~(\ref{LvsJ}) for $n=2$. \label{fig:Ra_vs_J}}
\end{figure*}

\section{Dynamic CTR evolution during coherent heteroepitaxial growth}

Real-time CTR measurements provide a powerful approach for probing epitaxial growth dynamics. However, extracting quantitative information remains challenging due to the coupled evolution of structural parameters, including film thickness, composition, and surface roughness. In particular, both the design of \textit{in situ} CTR experiments and the interpretation of measured intensity require a framework that directly links the scattering signal to the evolving atomic structure.

Building on the formalism developed in the previous section, we establish such a framework for coherent heteroepitaxial growth on vicinal surfaces. Rather than treating CTR intensity as a static function of reciprocal-space coordinates, the model naturally extends to a dynamic representation in the combined $(L, J)$ space. Figure~\ref{fig:Ra_vs_J} shows the calculated CTR intensity as a function of film thickness $J$ for various CTRs and $\alpha$-terrace fractions ($f_\alpha$), focusing on rods near the substrate Bragg peak $(H_0 K_0 L_0)$ with $L_0 = 2$. This representation reparameterizes conventional CTR profiles (e.g., Fig.~\ref{fig:CTR_off-cut2}), establishing a direct mapping between reciprocal-space scattering and growth evolution.

As the film grows, the CTR intensity exhibits oscillations arising from interference between the substrate and film scattering amplitudes, as shown in Fig.~\ref{fig:Ra_vs_J}. The phase of the substrate contribution changes by $\pi$ not only when crossing the Bragg peak, but also at intermediate positions where the scattering amplitude passes through minima. These minima are determined by the bulk GaN crystal structure (see Appendix~\ref{app:sub_minima}). In addition, due to terrace-dependent height offsets, the effective film thickness deviates from the layer index $J$. The average thickness is given by $J + f_\alpha/2$, reflecting the half-unit-cell height difference between $\alpha$ and $\beta$ terraces.

In the preceding formulation, the film thickness is expressed as $J$ (in u.c.). In the following, we use $d$ in monolayers (ML), with $J = d/2$ ($1~\mathrm{u.c.} = 2~\mathrm{ML}$).

To assess the feasibility of quantitative parameter extraction, simulated datasets are generated from the theoretical model with added Poisson noise. Unless otherwise specified, a growth rate of $0.1$ ML/s and a temporal resolution of $1$ s are assumed, representative of typical \textit{in situ} CTR measurements. Two measurement schemes are considered: $L$-scan during growth and real-time CTR at fixed scattering vector. These simulations are intended to establish sensitivity and fitting feasibility under controlled conditions, and do not aim to capture the full complexity of real growth processes.

\subsection{\label{sb:VA}\textit{In situ} evolution of CTR profiles along $L$}

\begin{figure}
\includegraphics[width=1\linewidth]{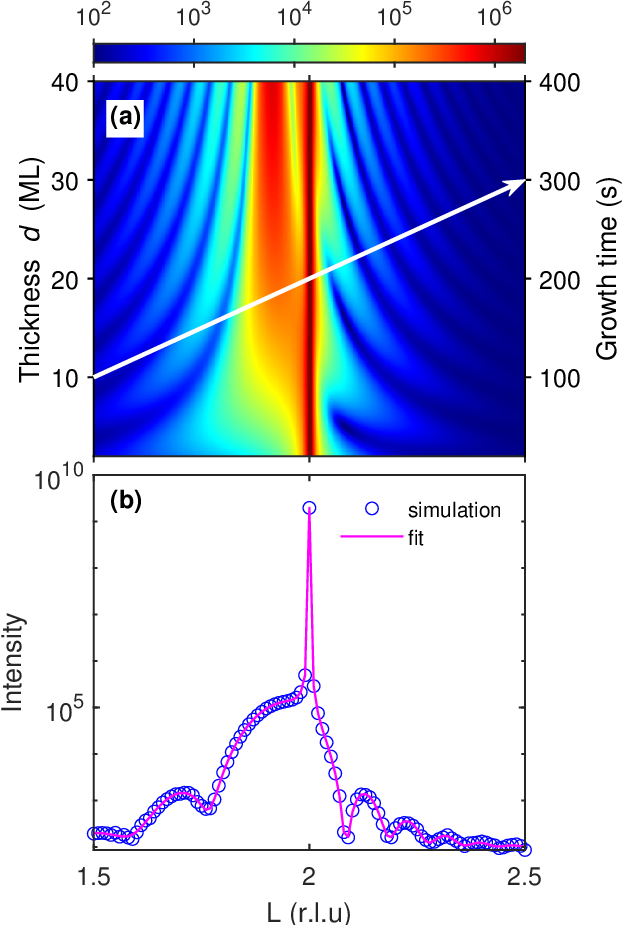}
\caption{Simulated CTR intensity during growth in the $(L, d)$ representation. 
(a) Color map of CTR intensity as a function of scattering coordinate $L$ and film thickness $d$. The white line indicates the measurement trajectory corresponding to an $L$-scan from $L = 1.5$ to $2.5$ while the film thickness $d$ increases from $ 10$ to $30$ ML. (b) Simulated $L$-scan profiles along this trajectory are shown together with the corresponding fits, where symbols denote simulated data with added Poisson noise. The extracted parameters are summarized in Table~\ref{tab:Lscan_simu}. This approach intrinsically couples reciprocal-space sampling with thickness evolution, thereby enabling direct and simultaneous extraction of structural parameters during growth.
\label{fig:Lscan_simu}}
\end{figure}

An effective approach for \textit{in situ} CTR measurements is to perform an $L$-scan during growth, in which $HK$ is fixed at $(H_0, K_0)$ while $L$ is continuously scanned along the CTR. In this configuration, the measured intensity does not correspond to a single film thickness, but instead traces a trajectory in the combined $(L, d)$ space as the film grows.

This measurement corresponds to a diagonal trajectory across the CTR intensity map in Fig.~\ref{fig:Lscan_simu}(a), intrinsically coupling reciprocal-space sampling with thickness evolution. As a result, each point in the measured intensity--$L$ curve corresponds to a different film thickness.
Compared with static CTR profiles at fixed thickness (Fig.~\ref{fig:CTR_off-cut2}), the overall line shape remains similar, but the interference fringes evolve continuously during the scan. In particular, as the film thickens, the fringe spacing decreases, leading to progressively denser oscillations at higher $L$, as illustrated in Fig.~\ref{fig:Lscan_simu}(b). This behavior reflects the continuous phase evolution of the CTR signal during growth, enabling the extraction of both structural and kinetic information from a single measurement.

Importantly, the fringe periodicity is highly sensitive to film thickness, which enables quantitative fitting of growth parameters such as thickness evolution, composition, and growth rate. This makes the $L$-scan approach particularly advantageous for real-time characterization of epitaxial growth.

To evaluate the feasibility of this approach, we simulate an $L$-scan measurement during growth under the conditions described above. An $L$ increment of $0.01$ r.l.u. is used for the scan. The fitting results are summarized in Table~\ref{tab:Lscan_simu}. Fitting error is calculated from the covariance matrix. Excellent agreement is obtained between the simulated and fitted parameters, including the indium composition, surface roughness, initial thickness, and growth rate. This demonstrates that the $L$-scan method enables accurate extraction of key growth parameters.

The $L$-scan approach provides improved stability and accuracy in determining structural parameters, owing to the strong sensitivity of fringe evolution to film thickness. 

\begin{table}[!]
\caption{\label{tab:Lscan_simu}
Comparison of simulated and fitted parameters for the $L$-scan during growth. $d_0$ denotes the film thickness at the start of the $L$-scan.}
\begin{ruledtabular}
\begin{tabular}{l l c c}
Parameter & Unit & Simulated & Fitted \\ \hline
$x_{\mathrm{In}}$ & — & 0.3 & 0.3003 $\pm$ 0.0004\\
$\sigma_R$ & (nm) & 0.2 & 0.2015 $\pm$ 0.0018\\
$d_0$ & (ML) & 10 & 10.072 $\pm$ 0.019\\
Growth rate & (ML/s) & 0.1 & 0.0987 $\pm$ 0.0003\\ \hline
\multicolumn{4}{l}{Best-fit $\chi^2_{\text{red}} = 1.15$}
\end{tabular}
\end{ruledtabular}
\end{table}

\subsection{\label{sb:fixedL}Real-time CTR at fixed $L$}

In addition to $L$-scan measurements, real-time CTR measurements can be performed at a fixed position $L$ along the CTR (or equivalently at a fixed scattering vector $\mathbf{q}$), where the scattered intensity is recorded as a function of growth time. In this configuration, the CTR intensity exhibits thickness-dependent oscillations with a period $\Delta J$ as the film grows, providing a direct probe of the phase evolution of the scattering signal. 

The oscillations originate from the thickness dependence of the total reflectivity amplitude $r_\mathrm{tot}$. The only terms containing the film thickness $J$ are those involving $Y_\mathrm{epi}^{JM}$, which appear in both $r_\mathrm{epi}$ and $r_\mathrm{rec}$. Since $\lvert Y_\mathrm{epi} \rvert \approx 1$, the oscillations are governed by the phase factor $\exp\left(i \mathbf{q} \cdot \mathbf{b}_\mathrm{epi} \right)^{JM}$. Evaluating $\mathbf{q} \cdot \mathbf{b}_\mathrm{epi}$ (see Appendix~\ref{app:iqb}) yields the corresponding phase evolution, from which the oscillation period (in u.c.) is obtained as
\begin{equation}
\Delta J = \frac{1}{\left|\frac{c_\mathrm{epi}}{c}L - L_0\right|} .
\label{eq:phaseJL}
\end{equation}
This relation shows that $\Delta J$ decreases as $L$ moves away from $L_0$, leading to increasingly rapid oscillations in the CTR intensity. Under fixed growth conditions and within the coherent model, $\Delta J$ is independent of thickness and is determined solely by the reciprocal-space coordinate $L$, and is therefore independent of the X-ray energy for fixed-$L$ measurements. 

The implications of this phase relation are illustrated in Fig.~\ref{fig:Ra_vs_J}, where the CTR intensity is shown as a function of 
$L$ and $J$. The oscillation period $\Delta J$ is independent of the $\alpha$-terrace fraction $f_{\alpha}$. The white curves correspond to trajectories of constant phase, defined by
\begin{equation}
L = \frac{c}{c_\mathrm{epi}}\left(L_0 \pm \frac{n}{J}\right),
\label{LvsJ}
\end{equation}
where $n$ is an integer indexing the oscillation order. For the case shown in Fig.~\ref{fig:Ra_vs_J}, $L_0 = 2$ and n=2 for all three CTR types of GaN. Along these trajectories, the phase of the interference term $Y_\mathrm{epi}^{JM}$ remains constant. Because this term carries the dominant thickness-dependent phase, these trajectories define the oscillatory structure of the CTR intensity in $(L,J)$ space.

The extrema of the reflectivity arise from the interference between the bulk and epilayer contributions. While the phase of $Y_\mathrm{epi}^{JM}$ sets the primary oscillation, the positions of the extrema relative to the equiphase trajectories are determined by the phase difference and magnitude ratio of the two contributions. Physically, the equiphase trajectories correspond to a fixed phase of the epilayer term, whereas the extrema result from constructive or destructive interference with the bulk. As both phase and amplitude evolve with $L$, their interplay shifts the extrema away from the equiphase condition. In geometric terms, the extrema correspond to points where $|r_\mathrm{tot}|$ is extremal in the complex plane (see Appendix~\ref{app:phase_extrema}). Thus, the equiphase trajectories generally follow the oscillatory pattern but do not coincide with the extrema, except when the bulk and epilayer contributions are excactly phase-aligned.

\begin{figure}[t!]
\includegraphics[width=0.9\linewidth]{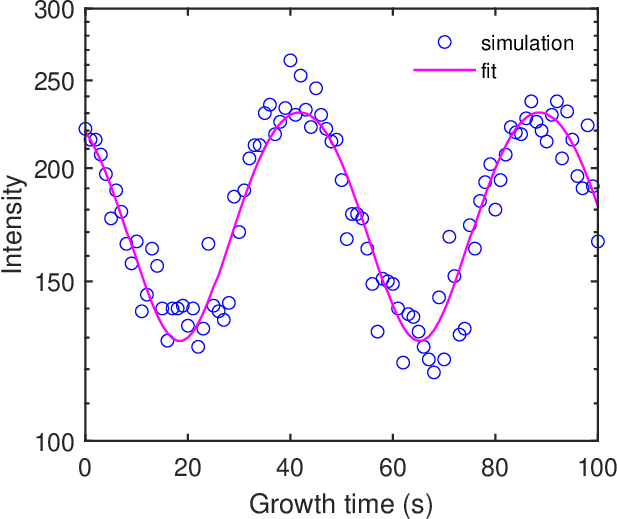}
\caption{Simulated time evolution of CTR intensity at fixed $L = 1.5$ on the (0002) CTR during growth. Symbols denote simulated data with Poisson noise; lines show the corresponding fits. The oscillatory behavior arises from interference between substrate and epilayer scattering amplitudes, with a period determined by Eq.~(\ref{eq:phaseJL}). The fitting results are summarized in Table~\ref{tab:fixedL_simu}. The agreement demonstrates that real-time CTR measurements at fixed $\mathbf{q}$ enable quantitative extraction of growth parameters, including composition, growth rate, and surface roughness.
\label{fig:fixedL_simu}}
\end{figure}

\begin{table}[b!]
\caption{\label{tab:fixedL_simu}
Comparison of simulated and fitted parameters for the fixed-$L$ CTR analysis.}
\begin{ruledtabular}
\begin{tabular}{l l c c}
Parameter & Unit & Simulated & Fitted \\ \hline
$x_{\mathrm{In}}$ & — & 0.3 & 0.281 $\pm$ 0.006\\
$\sigma_R$ & (nm) & 0.2 & 0.197 $\pm$ 0.003\\
Growth rate & (ML/s) & 0.1 & 0.0986 $\pm$ 0.0006\\ \hline
\multicolumn{4}{l}{$L = 1.5$ on (0002) CTR} \\
\multicolumn{4}{l}{Best-fit $\chi^2_{\text{red}} = 0.90$} \\
\multicolumn{4}{l}{$\Delta d$ = 4.69~ML $\approx$ 1.27~nm ($\Delta J$ = 2.34~u.c.)}
\end{tabular}
\end{ruledtabular}
\end{table}

From an experimental perspective, real-time CTR measurements record the scattered intensity as a function of growth time, from which the evolution of film thickness can be inferred. Different physical parameters affect distinct features of the intensity profile. Surface roughness primarily influences the overall intensity amplitude and background level, thereby determining the maximum and minimum intensities. In contrast, the growth rate and indium composition are intrinsically coupled [Eq.~(\ref{eq:phaseJL})] and jointly determine the oscillation period and the phase evolution of the CTR signal during growth..

To assess the feasibility of quantitative parameter extraction, we simulate real-time CTR measurements at fixed $L$ under the same conditions. The detector exposure time is set to correspond to a minimum resolvable thickness increment of approximately $0.1$ ML. In the data analysis, the overall intensity scale is assumed to be determined directly from the experimental data. In the fitting procedure, the growth rate $G$ is treated as an outer-loop parameter and scanned over a predefined range, while the remaining parameters are optimized within an inner fitting loop. The optimal values and their uncertainties are obtained from the minimum of the $\chi^2(G)$ curve.

The results are summarized in Table~\ref{tab:fixedL_simu}, showing good agreement between simulated and fitted parameters. This demonstrates that real-time CTR measurements at fixed $\mathbf{q}$ enable quantitative determination of structural parameters during growth. The intrinsic thickness sensitivity is determined by the oscillation period $\Delta d$, which reflects how rapidly the CTR phase evolves with film thickness and decreases as the selected $L$ position moves away from $L_0$ [see Eq.~(\ref{eq:phaseJL})]. A smaller $\Delta d$ improves intrinsic thickness sensitivity, but positions farther from $L_0$ generally provide weaker intensity and therefore require higher X-ray flux. As a result, the achievable thickness resolution reflects a trade-off between phase-based thickness sensitivity and signal-to-noise ratio.

Equation~(\ref{eq:phaseJL}) provides a direct route to determine the indium composition $x_{\mathrm{In}}$ without full curve fitting, provided that the growth rate is independently known (e.g., from Sec.~\ref{sb:VA} or \textit{in situ} laser reflectometry). The key quantity is the oscillation period $\Delta d$, which can be directly obtained from the temporal evolution of the CTR intensity. The oscillation period $\Delta d$ corresponds to the thickness increment over one full oscillation cycle and is therefore given by the product of the growth rate and the temporal oscillation period. This allows the lattice parameter ratio $c_\mathrm{epi}/c$ to be determined via Eq.~(\ref{eq:phaseJL}), with the conversion $\Delta J = \Delta d / 2$.

As an example, for a growth rate of $0.1$ ML/s, the measured oscillation period yields $\Delta d = 4.68$ ML (corresponding to $\Delta J$ = 2.34 u.c.). Substituting into Eq.~(\ref{eq:phaseJL}) with $L = 1.5$ and $L_0 = 2$ yields $c_\mathrm{epi}  = 1.048\,c$. According to Eq.~(\ref{eq:c_epi_N}), $c_\mathrm{epi}$ can be expressed as a monotonic function of the indium composition $x_{\mathrm{In}}$, with $a_\mathrm{epi}^0$ and $c_\mathrm{epi}^0$ determined via Vegard's law\cite{ju2013situ}. This enables a direct inversion to obtain $x_{\mathrm{In}}$ without fitting. In the present case, this procedure yields $x_{\mathrm{In}} = 0.305$, in good agreement with the nominal value despite the simplifying approximations underlying the model.

\section{Discussion}

The present work establishes a unified phase-based framework for understanding CTR scattering from vicinal surfaces in coherent heteroepitaxy.  Building on earlier treatments of vicinal surfaces without films\cite{2021_Ju_PRB_CTR}, the present formulation incorporates the lattice distortion of the epilayer, film-induced interference fringes, terrace-resolved surface structure, and real-time growth evolution within a unified scattering formalism. This directly links measured CTR profiles to the evolving atomic structure of a coherent heteroepitaxial film on a vicinal substrate.

A central result of the present work is the distinction between lattice tilt and triclinic deformation in coherent heteroepitaxy on vicinal surfaces. The Nagai model and the elasticity-based model yield nearly identical values of the lattice tilt angle $\delta$, showing that the tilt is largely governed by lattice mismatch and vicinal geometry. In contrast, the elasticity-based model predicts an additional triclinic deformation, characterized by the angle $\tau$, which originates from shear components of the strain tensor. The magnitude of $\tau$ is found to be comparable to, or even larger than, $\delta$, demonstrating that shear-induced distortion is an essential part of the elastic response. Although this distinction has little effect on specular CTRs, it produces pronounced changes in non-specular CTR profiles, where the scattering is directly sensitive to in-plane lattice distortion. This identifies non-specular CTRs as a particularly powerful probe of the full elastic state of coherently strained epitaxial films.

Another key finding is that the characteristic sensitivity of vicinal CTRs to terrace ordering and surface structure is preserved in the presence of a coherent film. Despite the additional complexity introduced by film-induced interference fringes, the dependence of CTR profiles on the $\alpha/\beta$ terrace fraction $f_\alpha$ remains robust. This demonstrates that step-resolved structural information can be extracted even under realistic heteroepitaxial conditions.

The formalism further enables the analysis of local compositional variations at the surface. By introducing terrace-resolved modified regions, the model captures step-selective incorporation processes and their signatures in CTR scattering. This provides a pathway to quantify local composition and segregation phenomena that are difficult to access by conventional \textit{in situ} techniques.

The real-time simulations further show how this framework can guide the design of \textit{in situ} growth experiments. For $L$-scan measurements during growth, the evolving fringe structure provides simultaneous sensitivity to film thickness, composition, and growth rate, enabling robust fitting of structural parameters from dynamically acquired CTR profiles. For measurements at fixed $L$ or $\mathbf{q}$, the analysis shows that the oscillation period is determined by the epilayer phase factor and depends on the selected $L$ position. In this case, the achievable thickness resolution is controlled by a trade-off between phase-based thickness sensitivity and signal-to-noise ratio: positions farther from $L_0$ give shorter oscillation periods and higher intrinsic thickness sensitivity, but at the cost of reduced scattered intensity. These results clarify how experimental conditions can be optimized depending on whether the primary goal is parameter extraction, high time resolution, or high thickness sensitivity.

The present framework is restricted to the coherent regime, where the terrace periodicity is inherited from the substrate and the film can be described by a single strained lattice. Within this assumption, effects such as plastic relaxation, step meandering, and disorder beyond the phenomenological roughness term are not included. The examples considered here further assume simplified interface structures and prescribed surface reconstructions. Extending the framework to incorporate more complex growth dynamics, nonequilibrium surface populations, and partially relaxed films remains an important direction for future work.

\section{Conclusions}

We have developed a general theoretical framework for crystal truncation rod scattering from vicinal surfaces with a coherently strained heteroepitaxial film. The formalism incorporates film-induced interference, full elastic lattice distortion, terrace-resolved surface structure, and real-time growth evolution within a unified description.

The results show that, while the lattice tilt predicted by the Nagai and elasticity-based models is nearly identical, the additional triclinic deformation arising from shear strain has a pronounced impact on non-specular CTRs, but remains negligible for specular rods. Non-specular CTRs therefore provide a direct probe of the full elastic state of coherent epitaxial films.

We further demonstrate that the characteristic sensitivity of vicinal CTRs to terrace ordering, surface reconstruction, and terrace-resolved compositional variations is preserved in the presence of a coherent film. The formalism also enables quantitative analysis of real-time CTR measurements, providing access to step-resolved structural and kinetic information during epitaxial growth.

Although the examples presented here focus on coherent InGaN/GaN heteroepitaxy, the formalism is general and can be extended to other vicinal heteroepitaxial systems. It establishes a quantitative and physically transparent basis for interpreting CTR scattering from vicinal heteroepitaxial systems and provides guidance for future \textit{in situ} experiments aimed at resolving atomic-scale growth dynamics.

\begin{acknowledgments}
This work was supported by the National Key Research and Development Program of China (Grant No. 2023YFE0124600) and the National Natural Science Foundation of China (Grant No. 62574008). The author gratefully acknowledges G. Brian Stephenson, Carol Thompson, Jeffrey A. Eastman, and Matthew J. Highland for valuable discussions and insightful input on CTR scattering, which greatly benefited this work.
\end{acknowledgments}

\appendix

\section{\label{sec:app.A}Elastic modeling and small-angle approximations for vicinal heteroepitaxy}

\subsection{Details of the elastic-based model}

The indices $1,2,3$ and $x,y,z$ are used interchangeably. All asymptotic orders below are taken with respect to the small off-cut angle $\theta$. 
The normal strains $\epsilon_{ii}$ are treated as finite parameters, while the shear components scale as $\epsilon_{23},\epsilon_{23}^\prime=\mathcal{O}(\theta)$.

The two coordinate systems mentioned in Sec.~\ref{sb:IIB} are related by
\begin{equation}
    \left[\begin{array}{l}
\hat{e}_x^{\prime} \\
\hat{e}_y^{\prime} \\
\hat{e}_z^{\prime}
\end{array}\right]=U^T\left[\begin{array}{l}
\hat{e}_x \\
\hat{e}_y \\
\hat{e}_z
\end{array}\right],
\end{equation}
where
\begin{equation}
    U^T=\left [ \begin{array}{ccc}
    1 & 0 & 0  \\
    0 & \cos(\theta_\mathrm{epi}^0) & -\sin(\theta_\mathrm{epi}^0)  \\
    0 & \sin(\theta_\mathrm{epi}^0) & \cos(\theta_\mathrm{epi}^0)
\end{array} \right ].
\end{equation}
Then 
\begin{align}    \boldsymbol{\epsilon^\prime}&=U^T\boldsymbol{\epsilon}U,\\
    \boldsymbol{\sigma^\prime}&=U^T\boldsymbol{\sigma}U.
    \label{aeq:coor_transf}
\end{align}
Hooke's law [Eq.~(\ref{eq:Hooke_law})] is written in Voigt notation as
\begin{equation}
    \left[\begin{array}{l}
\sigma_{11} \\
\sigma_{22} \\
\sigma_{33} \\
\sigma_{23} \\
\sigma_{31} \\
\sigma_{12}
\end{array}\right]=\left[\begin{array}{llllll}
c_{11} & c_{12} & c_{13} & & & \\
c_{12} & c_{11} & c_{13} & & & \\
c_{13} & c_{13} & c_{33} & & & \\
& & & c_{44} & & \\
& & & & c_{44} & \\
& & & & & \frac{c_{11}-c_{12}}{2}
\end{array}\right]\left[\begin{array}{c}
\epsilon_{11} \\
\epsilon_{22} \\
\epsilon_{33} \\
2\epsilon_{23} \\
2\epsilon_{31} \\
2\epsilon_{12}
\end{array}\right].
\label{aeq:hook}
\end{equation}

Together with Eq.~(\ref{eq:inplane}) and (\ref{eq:free_plane}),
\begin{align}
    \epsilon_{x^{\prime} x^\prime}=\epsilon_{m1};\quad \epsilon_{y^{\prime} y^{\prime}}&=\epsilon_{m2};\quad \epsilon_{x^{\prime} y^{\prime}}=0; \nonumber\\
    \sigma_{x^{\prime} z^{\prime}}=0;\quad \sigma_{y^{\prime} z^{\prime}}&=0;\quad \sigma_{z^{\prime} z^{\prime}}=0. \nonumber
\end{align}
Thus, the solution is uniquely determined. 

For later use, we summarize several relations:
\begin{align}
  \label{e23}  \epsilon_{23}&=\sigma_{23}/2c_{44}=-\sin\theta_\mathrm{epi}^0\cos\theta_\mathrm{epi}^0\sigma_{22}^\prime/2c_{44};\\
  \label{e23p}  \epsilon_{23}^\prime&=\sin\theta_\mathrm{epi}^0\cos\theta_\mathrm{epi}^0(\epsilon_{22}-\epsilon_{33})+\cos2\theta_\mathrm{epi}^0\epsilon_{23};\\  
  \label{e11p}  \epsilon_{11}^\prime&=\epsilon_{11};\\
  \label{e22p}  \epsilon_{22}^\prime&=(\cos\theta_\mathrm{epi}^0)^2\epsilon_{22}-\sin2\theta_\mathrm{epi}^0\epsilon_{23}+(\sin\theta_\mathrm{epi}^0)^2\epsilon_{33};\\
  \label{s33}  \sigma_{33}&=(\sin\theta_\mathrm{epi}^0)^2\sigma_{22}^\prime.
\end{align}

\subsection{\label{app:Small-quantity}Small-quantity expansion}

The purpose of this subsection is twofold: first, to show that the Nagai and elasticity-based models give the same lattice tilt to leading order; and second, to identify which components of the deformation matrix control the difference in CTR scattering. Here we assume that the off-cut angle $\theta \sim 1/M$ is small. In practice, $\theta$ usually does not exceed $2^\circ$. The linear strains $\epsilon_{ii}$ are usually less than 10\%, and the shear components $\epsilon_{23}^\prime$ and $\epsilon_{23}$ are of order $\epsilon_{ii}\theta$, as shown by Eq.~(\ref{e23}) and (\ref{e23p}). 

We first discuss the three angles $\gamma$, $\delta$, and $\tau$ in the elastic model. The rotation angle $\gamma$ is introduced so that the bottom surface of the strained epilayer remains parallel to the substrate surface. Equivalently, the rotated step-down direction satisfies the condition
\begin{equation}
    \boldsymbol{T}_x(\gamma)(\boldsymbol{I}+\boldsymbol{\epsilon}')\left[\begin{array}{l}
0 \\
1 \\
0 
\end{array}\right]\parallel\left[\begin{array}{l}
0 \\
\cos(\theta_\mathrm{epi}^0-\theta) \\
\sin(\theta_\mathrm{epi}^0-\theta)
\end{array}\right].
\label{aeq:Rx}
\end{equation}
This simplifies to
\begin{equation}
    \frac{\sin \gamma\left(1+\epsilon_{22}^{\prime}\right)+\cos \gamma \epsilon_{23}^{\prime}}{\cos \gamma\left(1+\epsilon_{22}^{\prime}\right)-\sin \gamma \epsilon_{23}^{\prime}}=\tan \left(\theta_\mathrm{epi}^0-\theta\right).
\end{equation}
Under the small-quantity approximation,
\begin{equation}
    \gamma=\theta_\mathrm{epi}^0-\theta+\frac{\epsilon_{23}^\prime}{1+\epsilon_{22}^\prime}+\mathcal{O}(\theta^3).
\end{equation}

The angle $\delta$ is the angle between the substrate $\mathbf{b}$ vector and the corresponding $\mathbf{b}_\mathrm{epi}$ vector in the strained epilayer. It describes the lattice tilt of the epilayer. Geometrically,
\begin{equation}
    \delta=\gamma+\arctan[\epsilon_{23}/(1+\epsilon_{22})].
    \label{eq:del}
\end{equation}

The angle $\tau$ is defined as the complementary angle to the angle between the lattice vector $\mathbf{c}_\mathrm{epi}$ and $\mathbf{b}_\mathrm{epi}$. Including the shear strain ($\boldsymbol{I+\epsilon}$), one obtains
\begin{equation}
\tau=\arctan[\epsilon_{23}/(1+\epsilon_{33})]+\arctan[\epsilon_{23}/(1+\epsilon_{22})].
\end{equation}
In most cases, the biaxial strain $\epsilon_{ii}$ is less than 10\%, while $\epsilon_{22}$ and $\epsilon_{33}$ have opposite signs, so that $\tau\approx2\epsilon_{23}$ 
, which leads to
\begin{equation}
    \delta\approx\gamma+(\tau/2).
\end{equation}

Thus, $\gamma$, $\delta$, and $\tau$ are all small angles of order $\mathcal{O}(\theta)$, with the shear-induced part scaling as $\mathcal{O}(\epsilon_{ii}\theta)$.

In Sec.~\ref{sec:II}, we noted that the two crystal models yield very similar values of $\delta$.  For Nagai's model, 
\begin{equation}
    \tan\delta_N=\frac{c_{\mathrm{epi},N}-c}{Mb},
    \label{eq:dN}
\end{equation}
whereas for the elastic model,
\begin{align}
\tan\delta_e&=\frac{(1+\epsilon_{33})c_\mathrm{epi}^0\cos(\gamma+\epsilon_{23}/(1+\epsilon_{33}))-c}{Mb+(1+\epsilon_{33})c_\mathrm{epi}^0\sin(\gamma+\epsilon_{23}/(1+\epsilon_{33}))}\nonumber\\
&=\frac{(1+\epsilon_{33})c_\mathrm{epi}^0-c}{Mb}+\mathcal{O}(\theta^2).
    \label{eq:de}
\end{align}
From Hooke's law 
\begin{equation}
    c_{33}\epsilon_{33} = -c_{13}\epsilon_{11}-c_{13}\epsilon_{22}+\sigma_{33}.
    \label{eq:e33}
\end{equation}
Since $\boldsymbol{\epsilon^\prime}$ and $\boldsymbol{\epsilon}$ are related by $U$, one has
\begin{equation}
\epsilon_{11}=\epsilon_{m1};\epsilon_{22}=\epsilon_{m2}+\mathcal{O}(\theta^2);\sigma_{33}=\mathcal{O}(\theta^2);
\label{eq:e11=em1}
\end{equation}
using Eqs.~(\ref{e11p}), (\ref{e22p}), and (\ref{s33}). In addition, 
\begin{equation}
\epsilon_{m2}=\frac{\sqrt{(Mb)^2 + c^2}}{\sqrt{(Mb_\mathrm{epi}^0)^2 + (c_\mathrm{epi}^0)^2}}-1=\frac{b}{b_\mathrm{epi}^0}-1+\mathcal{O}(\theta^2),
\label{eq:e22_to_b_over_b0}
\end{equation}
substituting these relations into Eq.~(\ref{eq:e33}) gives
\begin{align}
    c_\mathrm{epi}^0(1+\epsilon_{33})&=c_\mathrm{epi}^0(1-\frac{c_{13}}{c_{33}} 
    \frac{a-a_\mathrm{epi}^0}{a_\mathrm{epi}^0} -\frac{c_{23}}{c_{33}} 
    \frac{b-b_\mathrm{epi}^0}{b_\mathrm{epi}^0})+\mathcal{O}(\theta^2)\nonumber\\
    &=c_{\mathrm{epi},N}+\mathcal{O}(\theta^2),
    \label{eq:e33 to c/c0}
\end{align}
where Eq.~(\ref{eq:strain_Nagai}) has been used. Substituting Eq.~(\ref{eq:dN}) and (\ref{eq:de}) then yields
\begin{equation}
    \delta_N=\delta_e+\mathcal{O}(\theta^2).
\label{eq:delta}
\end{equation}

In Sec.~\ref{sb:IIIB}, we noted that the difference in CTR intensity between two crystal models is negligible in the specular case but more significant in the non-specular case. 
The underlying reason is that the scalar product $\mathbf{q}\cdot(\boldsymbol{\Delta^\prime_e}-\boldsymbol{\Delta^\prime_N})$ is close to zero when $q_y=2\pi K/b=0$. 
For Nagai's model, 
\begin{equation}
    \boldsymbol{\Delta^\prime_N}=\boldsymbol{T}_x(\delta)\left [ \begin{array}{ccc}
    a_\mathrm{epi}/a_\mathrm{epi}^0 & 0 & 0  \\
    0 & b_\mathrm{epi}/b_\mathrm{epi}^0 & 0  \\
    0 & 0 & c_\mathrm{epi}/c_\mathrm{epi}^0
\end{array} \right ],
\end{equation}
whereas for the elastic model, 
\begin{equation}
    \boldsymbol{\Delta^\prime_e}=\boldsymbol{T}_x(\gamma)\left [ \begin{array}{ccc}
    1+\epsilon_{11} & 0 & 0  \\
    0 & 1+\epsilon_{22} & \epsilon_{23}  \\
    0 & \epsilon_{23} & 1+\epsilon_{33}
\end{array} \right ].
\end{equation}
We therefore evaluate $\boldsymbol{\Delta^\prime_e}-\boldsymbol{\Delta^\prime_N}$ element by element.
For the $(1,1)$  element, (\ref{eq:e11=em1}) gives
\begin{align}
    ( \boldsymbol{\Delta^\prime_e}-\boldsymbol{\Delta^\prime_N})_{11}=
    1+\epsilon_{11}-\frac{a_\mathrm{epi}}{a_\mathrm{epi}^0}=0.
\end{align}
Thus, the two models give identical contributions to the specular $(H00)$ CTR at this order.

For the $(3,2)$  element, 
\begin{align}
    &( \boldsymbol{\Delta^\prime_e}-\boldsymbol{\Delta^\prime_N})_{32}=
    (1+\epsilon_{22})\sin\gamma+\epsilon_{23}\cos\gamma- \frac{b_\mathrm{epi}}{b_\mathrm{epi}^0}\sin\delta\nonumber\\
    &=(1+\epsilon_{22})\left(\sin\gamma+\frac{\epsilon_{23}}{1+\epsilon_{22}}\cos\gamma- \sin\delta\right)+\mathcal{O}(\theta^2)\nonumber\\
    &=\mathcal{O}(\theta^2),
\end{align}
using Eqs.~(\ref{e22p}) and (\ref{eq:e22_to_b_over_b0}), together with Eq.~(\ref{eq:del}).

For the $(3,3)$  element, 
\begin{align}
    &( \boldsymbol{\Delta^\prime_e}-\boldsymbol{\Delta^\prime_N})_{33}
    =\epsilon_{23}\sin\gamma+(1+\epsilon_{33})\cos\gamma-\frac{c_\mathrm{epi}}{c_\mathrm{epi}^0}\cos\delta\nonumber\\
    &=
    (1+\epsilon_{33})-\frac{c_\mathrm{epi}}{c_\mathrm{epi}^0}+\mathcal{O}(\theta^2)=\mathcal{O}(\theta^2),
\end{align}
using Eq.~(\ref{eq:e33 to c/c0}). 
For the $(2,2)$ element, 
\begin{align}
    &( \boldsymbol{\Delta^\prime_e}-\boldsymbol{\Delta^\prime_N})_{22}
    =(1+\epsilon_{22})\cos\gamma-\epsilon_{23}\sin\gamma- \frac{b_\mathrm{epi}}{b_\mathrm{epi}^0}\cos\delta\nonumber\\
    &=(1+\epsilon_{22})-\frac{b_\mathrm{epi}}{b_\mathrm{epi}^0}+\mathcal{O}(\theta^2)=\mathcal{O}(\theta^2),
\end{align}
using Eq.~(\ref{eq:e22_to_b_over_b0}). 
For the $(2,3)$ element, it is
\begin{align}
    &( \boldsymbol{\Delta^\prime_e}-\boldsymbol{\Delta^\prime_N})_{23}
    =\epsilon_{23}\cos\gamma-(1+\epsilon_{33})\sin\gamma+\frac{c_\mathrm{epi}}{c_\mathrm{epi}^0}\sin\delta\nonumber\\
    &=\epsilon_{23}+(1+\epsilon_{33})(\delta-\gamma) +\mathcal{O}(\theta^3)\nonumber\\
    &=\epsilon_{23}+(1+\epsilon_{33})\frac{\epsilon_{23}}{1+\epsilon_{22}} +\mathcal{O}(\theta^3)\nonumber\\
    &\approx 2\epsilon_{23}=\mathcal{O}(\theta),
\end{align}
using Eq.~(\ref{eq:del}). 

Among the matrix elements of $\boldsymbol{\Delta}_e' - \boldsymbol{\Delta}_N'$, the leading-order contribution is the $(2,3)$ component, which is of order $\mathcal{O}(\theta)$, whereas all other nonzero elements are of order $\mathcal{O}(\theta^2)$. Consequently, the scalar product $\mathbf{q}\cdot(\boldsymbol{\Delta}'_e-\boldsymbol{\Delta}'_N)$ remains negligible when $q_y=2\pi K/b=0$, i.e., along specular CTRs. This explains why the two models yield nearly identical intensities for specular CTRs, while more pronounced differences arise for non-specular CTRs, where sensitivity to shear-related distortions becomes significant.

\section{Phase-based analysis of CTR oscillations in coherent heteroepitaxy}
For all CTR-related variables in this section, we use the same parameters as in Sec.~\ref{sb:fixedL}.

\subsection{\label{app:iqb}Phase accumulation and thickness oscillation period}

We first analyze the dependence of the CTR phase on the film thickness $J$. 
In the expression for the total reflectivity amplitude $r_{\mathrm{tot}}$, the only term that depends on $J$ is $Y_{\mathrm{epi}}^{JM}$, which appears in both $r_{\mathrm{epi}}$ and $r_{\mathrm{rec}}$. 
Since $|Y_{\mathrm{epi}}| \approx 1$, its magnitude deviates only weakly from unity, and the oscillatory behavior is governed primarily by the phase factor $\exp\!\left(i \mathbf{q} \cdot \mathbf{b}_{\mathrm{epi}}\right)^{JM}$. 
We therefore evaluate $\mathbf{q} \cdot \mathbf{b}_{\mathrm{epi}}$.

For a vicinal geometry, the vector $\mathbf{b}_{\mathrm{epi}}$ lies in the $yz$ plane. Using the geometric relation for the step-down direction in the strained lattice, it can be decomposed into its components along the $y$ and $z$ directions as
\begin{align}
\mathbf{q}\cdot \mathbf{b}_\mathrm{epi}
&= q_y [b+\frac{c_\mathrm{epi}}{M}\sin(\tau-\delta)] + q_z\frac{c_\mathrm{epi}-c}{M}\cos(\tau-\delta)\nonumber\\
&= q_y b + q_z\frac{c_\mathrm{epi}-c}{M} + \mathcal{O}(\theta^2),
\end{align}
where we have used the fact that $1/M$, $\delta$, and $\tau$ are all of order $\theta$, so that $\sin(\tau-\delta)=\mathcal{O}(\theta)$ and $\cos(\tau-\delta)=1+\mathcal{O}(\theta^2)$.

Substituting $q_y = 2\pi K/b$ and $q_z = 2\pi L/c$, and using the relation between $K$ and $L$ along the CTR [Eq.~(\ref{eq:K})], we obtain
\begin{align}
\mathbf{q}\cdot \mathbf{b}_\mathrm{epi}
&= 2\pi\left[ K_0 + \frac{1}{M}\left( \frac{c_\mathrm{epi}}{c}L - L_0 \right) \right] + \mathcal{O}(\theta^2).
\end{align}

Thus, to leading order, the phase accumulation is governed by the mismatch in the effective out-of-plane lattice parameter. Removing the integer multiple $2\pi K_0$ and restricting the phase to the interval $(-\pi,\pi]$, the oscillation period is obtained as Eq.~(\ref{eq:phaseJL}).

\subsection{\label{app:phase_extrema} Complex-plane interpretation of equiphase trajectories and reflectivity extrema}

In Sec.~\ref{sb:fixedL}, we noted that the trajectories of constant phase lie close to the extrema of the reflectivity. To clarify the origin of this behavior, we consider a simplified case in which surface reconstruction is neglected. This behavior can be understood geometrically in the complex plane. For a representative example with $L_0=2$ and $L=1.5$, the total reflectivity can be written as
\begin{equation}
r_\mathrm{tot}=r_\mathrm{bulk}+z_\mathrm{epi}\left(1-Y_\mathrm{epi}^{JM}\right),
\end{equation}
with
\begin{align}
r_\mathrm{bulk}&=\frac{r_\mathrm{f}}{M}F_\mathrm{bulk}\frac{Y_\mathrm{bulk}}{Y_\mathrm{bulk}-1},\nonumber\\
z_\mathrm{epi}&=\frac{r_\mathrm{f}}{M}F_\mathrm{epi}\frac{Y_\mathrm{epi}}{Y_\mathrm{epi}-1},
\end{align}
where $z_\mathrm{epi}$ is a thickness-independent complex prefactor.

\begin{figure}[t]
\includegraphics[width=1\linewidth]{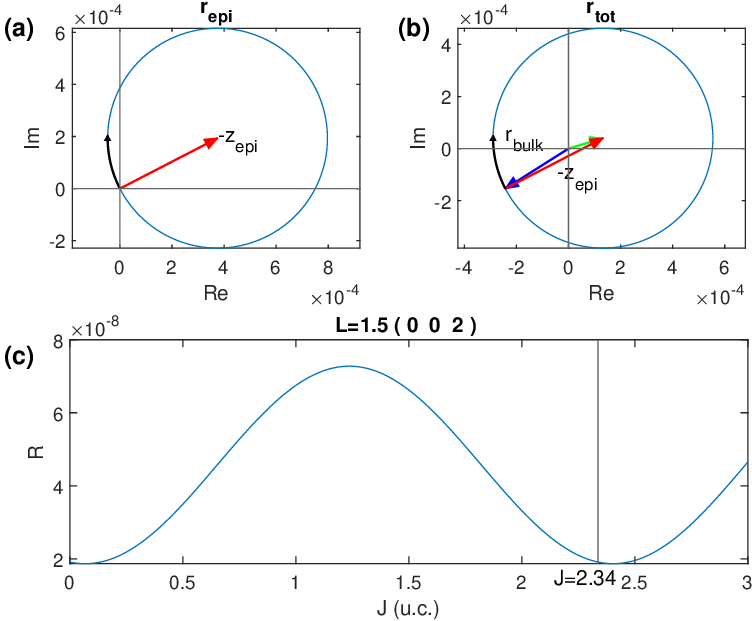}
\caption{Complex-plane trajectory of (a) $r_\mathrm{epi}$ and (b) $r_\mathrm{tot}$ (excludes surface reconstruction), and (c) reflectivity $R$ vs. thickness $J$, for the (002) CTR at fixed $L=1.5$. The blue vector: $r_\mathrm{bulk}$; red: $-z_\mathrm{epi}$; green: $r_\mathrm{bulk}-z_\mathrm{epi}$ pointing to center of $r_\mathrm{tot}$ circle. The black arrow on circle indicates increasing $J$ from given equiphase $J=n\Delta J$.
\label{fig:Rphase}}
\end{figure}

As the film thickness $J$ increases, the factor $Y_\mathrm{epi}^{JM}$ traces out a unit circle in the complex plane, since $|Y_\mathrm{epi}|\approx 1$. Its phase evolves as $2\pi J/\Delta J$, corresponding to uniform motion along the circle. Consequently, the quantity $1-Y_\mathrm{epi}^{JM}$ traces a circle centered at $(-1,0)$, and the epilayer contribution
\begin{equation}
r_\mathrm{epi}=z_\mathrm{epi}\left(1-Y_\mathrm{epi}^{JM}\right)
\end{equation}
describes a circle passing through the origin with its center at $-z_\mathrm{epi}$, as shown in Fig.~\ref{fig:Rphase}(a).

%(a) Complex-plane trajectory of $r_\mathrm{epi}$, (b) complex-plane trajectory of $r_\mathrm{tot}$, and (c) corresponding reflectivity $R$ as a function of film thickness $J$, for the (002) CTR at fixed $L=1.5$. $r_\mathrm{tot}$ here does not include surface reconstruction. The blue vector represents $r_\mathrm{bulk}$, and the red vector represents $-z_\mathrm{epi}$. The green vector $r_\mathrm{bulk}-z_\mathrm{epi}$ points to the center of the $r_\mathrm{tot}$ circle. The black arrow on the circle indicates the position corresponding to a given equiphase $J$ and the direction of evolution of $r_\mathrm{epi}$ and $r_\mathrm{tot}$ as $J$ increases.

The total reflectivity $r_\mathrm{tot}$ is obtained by translating this circle by $r_\mathrm{bulk}$, so that $r_\mathrm{tot}$ traces a circle centered at $r_\mathrm{bulk}-z_\mathrm{epi}$, as illustrated in Fig.~\ref{fig:Rphase}(b). As $J$ increases, $r_\mathrm{tot}$ moves along this trajectory, and the reflectivity $R \propto |r_\mathrm{tot}|^2$ attains extrema when $r_\mathrm{tot}$ lies along the line passing through the origin and the circle center $r_\mathrm{bulk}-z_\mathrm{epi}$.

Trajectories of constant phase, $J=n\Delta J$, as determined by Eq.~(\ref{LvsJ}), correspond to fixed angular position on $r_\mathrm{tot}$ circle. The position is exactly the point $r_\mathrm{bulk}$ corresponding to $r_\mathrm{tot}\rvert_{J=0}$. In the present system, $r_\mathrm{bulk}$ happens to lie close to the line passing through origin and $r_\mathrm{bulk}-z_\mathrm{epi}$, but on the opposite side of the origin, so that the corresponding equiphase trajectories are located near reflectivity minima. In the special case where $r_\mathrm{bulk}-z_\mathrm{epi}$ is parallel to $r_\mathrm{bulk}$, the equiphase trajectories would coincide exactly with the extrema of $R$.

This near coincidence originates from the similar phases of the structure factors $F_\mathrm{bulk}$ and $F_\mathrm{epi}$. For example, for $HKL=(0,0,1.5)$, their phases are $0.58$ and $0.49$(rad), respectively, reflecting the similar atomic configurations of the GaN substrate and the InGaN epilayer. Since $Y_\mathrm{bulk}, Y_\mathrm{epi} \approx 1$, the factor $Y/(Y-1)$ amplifies the difference in magnitude between the bulk and epilayer contributions, while preserving their similar phase. As a result, $r_\mathrm{bulk}$ lies close to the extremal direction of the $r_\mathrm{tot}$ circle, leading to the observed proximity between equiphase trajectories and reflectivity extrema. This behavior is consistently observed for different CTRs in the present system (see the white curves in Fig.~\ref{fig:Ra_vs_J}), but is not expected to be universal.  In general heteroepitaxial systems, the structure factors of the substrate and epilayer may differ significantly in phase, in which case the alignment between $r_\mathrm{bulk}$ and $r_\mathrm{bulk}-z_\mathrm{epi}$ is lost, and the equiphase trajectories no longer track the reflectivity extrema.

\subsection{\label{app:sub_minima}Phase jumps and extrema switching near substrate minima}

We observe a rapid phase shift along the equiphase trajectories near minima of the total reflectivity $R(L)$. This behavior is particularly evident near $L=1$ and $L=3$ in Fig.~\ref{fig:Ra_vs_J} for the (00L) CTR with $f_\alpha=0$ and $f_\alpha=1$, where the trajectories that initially correspond to reflectivity minima switch to maxima, and vice versa.

\begin{figure*}
\includegraphics[width=0.8\linewidth]{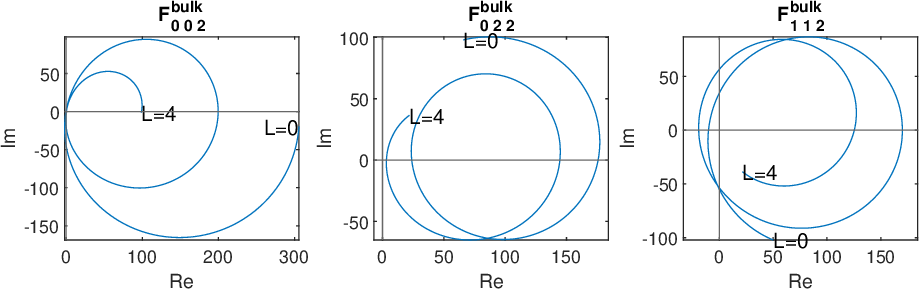}
\caption{Complex-plane trajectories of the bulk structure factor $F_\mathrm{bulk}$ for the (002), (022), and (112) CTRs of a vicinal GaN$(0001)$ surface. As $L$ varies, $F_\mathrm{bulk}$ traces spiral-like paths. Points at $L=0$ and $L=4$ are marked. The approach to or crossing of the origin corresponds to rapid phase variation or $\pi$ phase jumps.%, which play a key role in the observed phase behavior of the CTR intensity.
\label{fig:Fspiral}}
\end{figure*}

A natural interpretation is to attribute this phase shift solely to the substrate contribution. The bulk reflectivity $r_\mathrm{bulk}(L)$ can be decomposed into the structure factor $F_\mathrm{bulk}$ and the geometric factor $Y_\mathrm{bulk}/(Y_\mathrm{bulk}-1)$. The latter dominates near Bragg positions, where it diverges and produces a $\pi$ phase jump. In addition, the phase of $F_\mathrm{bulk}$ evolves with $L$ and traces an irregular spiral in the complex plane (Fig.~\ref{fig:Fspiral}). When $F_\mathrm{bulk}$ approaches or crosses the origin, the phase changes rapidly or undergoes a $\pi$ jump, as observed at $L=1$ and $L=3$ for the (002) CTR. However, this substrate-only picture is insufficient to explain the observed switching behavior. In particular, for the (00L) CTR with $f_\alpha=0$, the phase shift near $L\approx 3$ occurs slightly below the nominal extinction position, indicating an additional contribution beyond the substrate phase jump.

The key point is that the epilayer contribution $z_\mathrm{epi}$ exhibits a similar phase evolution to $r_\mathrm{bulk}$, as shown in Fig.~\ref{fig:FFf}. As established in Appendix~\ref{app:phase_extrema}, the relative location of equiphase trajectories with respect to reflectivity extrema is determined not only by the phase difference between $r_\mathrm{bulk}$ and $z_\mathrm{epi}$, but also by their relative magnitudes.

\begin{figure*}
\includegraphics[width=0.8\linewidth]{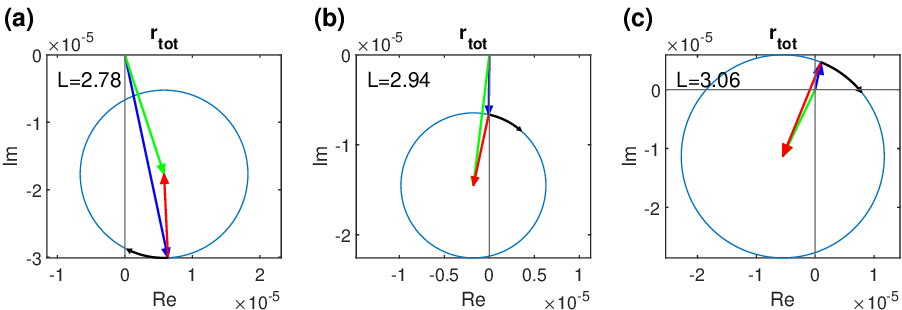}
\caption{Complex-plane trajectories of $r_\mathrm{tot}$ for the (002) CTR at $L=2.78$, $2.94$, and $3.06$. The total reflectivity does not include surface reconstruction. The blue and red vectors represent $r_\mathrm{bulk}$ and $-z_\mathrm{epi}$, respectively, and the green vector $r_\mathrm{bulk}-z_\mathrm{epi}$ points to the circle center. The black arrow indicates the equiphase position and the direction of evolution with increasing $J$. The change in the relative orientation of these vectors illustrates the switching of reflectivity extrema.
\label{fig:rtot_circle}}
\end{figure*}

Specifically, for $2<L<2.87$, the phases of $r_\mathrm{bulk}$ and $z_\mathrm{epi}$ are nearly identical, while $|r_\mathrm{bulk}|>|z_\mathrm{epi}|$. In this regime, the bulk term dominates, and the equiphase trajectories lie near reflectivity maxima [Fig.~\ref{fig:rtot_circle}(a)]. As $L$ increases to $2.87<L<3$, the phase difference approaches $\pi$ and $|r_\mathrm{bulk}|<|z_\mathrm{epi}|$, so that the phase difference becomes dominant and the equiphase trajectories shift toward reflectivity minima [Fig.~\ref{fig:rtot_circle}(b)]. For $L>3$, the phase difference becomes small again, but the magnitude relation $|r_\mathrm{bulk}|<|z_\mathrm{epi}|$ persists, and the trajectories remain near minima [Fig.~\ref{fig:rtot_circle}(c)]. These results demonstrate that the phase shift and extrema switching arise from the combined effect of phase difference and modulus competition between the bulk and epilayer contributions, rather than from the substrate phase alone.

\begin{figure*}
\includegraphics[width=0.8\linewidth]{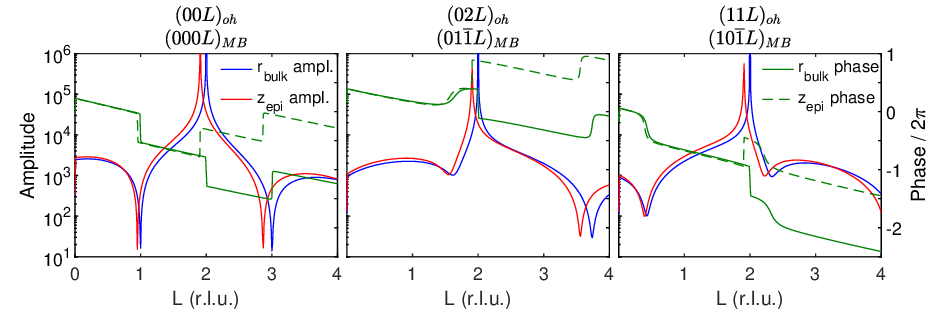}
\caption{Calculated amplitude and phase of the bulk and epilayer contributions for the $(00L_0)$, $(02L_0)$, and $(11L_0)$ CTRs at $L_0=2$, for a vicinal GaN $(0001)$ surface with a coherently strained In$_{0.3}$Ga$_{0.7}$N epitaxial film. Blue and red curves denote the amplitudes of $r_\mathrm{bulk}$ and $z_\mathrm{epi}$, respectively (left axis), while green solid and dashed curves show their corresponding phases (right axis). The rapid phase variation near amplitude minima and Bragg positions highlights the role of phase evolution in determining the interference behavior of the CTR intensity. 
\label{fig:FFf}}
\end{figure*}

\clearpage
\bibliography{2022_GJu_ABSteps_full_shortJnames}

\end{document}